\documentclass[twocolumn]{aastex63}

\usepackage{xcolor}
\usepackage{natbib}
\usepackage{graphicx}
\usepackage{amssymb}
\usepackage{multirow}
\usepackage{wrapfig}
\usepackage{enumitem}
\usepackage[varg]{txfonts}
\usepackage[toc,title,page]{appendix}
\usepackage{csquotes}

\def\baselinestretch{1}
\bibpunct{(}{)}{;}{a}{}{,}

\usepackage{hyperref}

\newcommand*{\kms}{km s$^{-1}$}

\newcommand*{\z}{$|Z|$}
\newcommand*{\I}{$(2I_3)^{1/2}$}

\newcommand*{\kpckms}{kpc~km s$^{-1}$}

\newcommand*{\km}{km$^{2}$s$^{-2}$}
\newcommand{\bmf}[1]{\mbox{{\it \boldmath$#1$}}}

\definecolor{tropicalrainforest}{rgb}{0.0, 0.46, 0.37}

\hypersetup{
    bookmarks=true,         
    unicode=false,          
    pdftoolbar=true,        
    pdfmenubar=true,        
    pdffitwindow=true,     
    pdfstartview={FitH},    
    pdftitle={},    
    pdfauthor={Beers},     
    pdfsubject={Astronomy},   
    pdfcreator={dvipdf},   
    pdfproducer={dvipdf}, 
    pdfkeywords={metal-poor stars},
    pdfnewwindow=true,      
    colorlinks=true,       
    linkcolor=red,          
    citecolor=blue,        
    filecolor=magenta,      
    urlcolor=cyan,           
    breaklinks=true,
    linktocpage
}

\begin{document}
\title{The nature of the Milky Way's stellar halo revealed by the three integrals of motion}

\author{Daniela Carollo}
\affiliation{INAF - Osservatorio Astronomico di Trieste, I-34143 Trieste, Italy}

\author{Masashi Chiba}
\affiliation{Astronomical Institute, Tohoku University, Sendai 980-8578, Japan}

\begin{abstract}
We developed a new selection method of halo stars in the phase-space distribution defined by the three integrals of motion in an axisymmetric Galactic potential, ($E$, $L_z$, $I_3$), where $I_3$ is the third integral of motion. The method is used to explore the general chemo-dynamical structure of the halo based on stellar samples from SDSS-SEGUE DR7 and DR16-APOGEE, matched with Gaia-DR2. We found, (a) halo stars can be separated from disk stars by selecting over (1) $0 < L_z < 1500$ \kpckms, $(2I_3)^{1/2} > 1000$ \kpckms (orbital angle $\theta_{\rm orb}$ $>$ 15-20 deg), and $E < -1.5 \times 10^5$ km$^2$ s$^{-2}$, and (2) $L_z < 0$ \kpckms. These selection criteria are free from kinematical biases introduced by the simple high-velocity cuts adopted in recent literature; (b)  the averaged, or {\it coarse-grained}, halo phased-space distribution shows a monotonic exponential decrease with increasing $E$ and $I_3$ like the Michie-Bodenheimer models; (c) the inner stellar halo described in \citet{carollo2007,carollo2010} is found to comprise a combination of Gaia Enceladus debris (GE; \citealt{helmi2018}), lowest-$E$ stars (likely in-situ stars), as well as metal-poor prograde stars missed by the high velocity cuts selection; (d) the very metal poor outer halo, ([Fe/H] $< -$2.2), exhibits both retrograde and prograde rotation, with an asymmetric $L_z$ distribution towards high retrograde motions, and larger $\theta_{\rm orb}$ than those possessed by the GE dominated inner halo; (e) the Sgr dSph galaxy could induce a long-range dynamical effect on local halo stars. Implication for the formation of the stellar halo are also discussed.
\end{abstract}


\section{Introduction}

The MW's halo is a gold mine of information on the assembly process and chemical enrichment that led to the Galaxy
we observe today. 
Since \citet{els1962} first presented their view of Galaxy formation based on the analysis of nearby
stars with various metallicities and orbits, our picture of how MW-like galaxies formed has developed and evolved
significantly, initially driven by the advent of newly calibrated data sets
\citep{sz1978,yoshii1979,norris1985,beers1995,carney1996}. 
Based on the properties of a kinematically biased sample of halo stars, in which metal-poor stars showed predominantly high orbital eccentricities ($e$), Eggen et al. proposed a monolithic, infalling and dissipative collapse scenario of a proto-Galactic
gas cloud for the formation of MW.
Subsequent works, where non-kinematically selected halo samples were adopted, showed the existence of low-$e$ metal-poor stars,
and significantly challenged the monolitic collapse model \citep{yoshii1979,norris1985}.

Modern stellar catalogs, such as those based on the first astrometric satellite, {\it Hipparcos}, provided further
evidence for the existence of low-$e$ halo stars \citep{chiba1998}, and even showed the first signature of the hierarchical assembly of
the stellar halo \citep{helmi1999}, in agreement with the seminal work by \citet{sz1978}, who proposed that the stellar
halo formed through dissipantion-less chaotic merging and accretion of many subsystems.
In fact, both dissipative and dissipation-less processes appear to be at work in the formation of the stellar halo,
as initially shown by the comparison between the observed orbital properties of halo stars and the predictions of CDM-based galaxy formation models \citep{bekki2001}.
The nature and complexity of the MW's halo was further unfolded
by subsequent studies based on the analysis of large stellar samples, in particular, its multiple nature and the presence of numerous individual streams and over-densities \citep[see reviews by, e.g.,][]{freeman2002,helmi2008,ivezic2012,feltzing2013,bland2016}.


Evidence that the halo may comprise more than one stellar population was coming from the analysis of its spatial profile
\citep{sommer1990,preston1991,zinn1993,kinman1994,miceli2008}, and the indication of retrograde motion of halo stars
\citep{majewski1992,carney1996,wilhelm1996,kinman2007,lee2007}. However, the first clear demonstration that the Milky Way's
halo comprises \enquote{at least} two stellar populations with different kinematics, spatial distribution, and
chemical composition was described in \citet{carollo2007,carollo2010} (see also \citealt{beers2012}).

The second release of the Gaia mission, {\it Gaia} DR2 \citep{brown2018}, has provided a large data set with unprecedented
high precision astrometric parameters, and a significant number of works made use of these data, in combination with radial
velocities and stellar abundances, tackling every component of the Milky Way. 
These works have revealed an even more complex but detailed picture of the Galaxy's halo. For instance, \citet{helmi2018}
\citep[see also][]{belokurov2018} showed that, at [Fe/H] $< -1$, the local halo is dominated by debris stars product of
a merging event occurred $\sim$ 10 Gyr ago (Gaia Enceladus; GE).
Several other works have reported the presence of halo streams possessing both prograde and retrograde motion, relative to
 the Galactic disk rotation direction, and also found new chemo-dynamical properties of the stellar halo,
that were unknown before {\it Gaia} \citep[e.g.,][]{bonaca2017,myeong2018a,myeong2018b,myeong2018c,myeong2018d,koppelman2019, matsuno2019, belokurov2020, yuan2020, bonaca2020,naidu2020}.

For the correct understanding of the stellar halo and its formation scenario, it is crucial to select halo stars without
introducing any biases, while reducing the contamination from the disk components. Such a selection can be partly
accomplished by adopting high velocity cuts because halo stars move differently from the Sun,
and this selection method has frequently been adopted in recent works \citep{nissen2010,bonaca2017,helmi2018}. 
However, we emphasize that important parts of the stellar halo are missed by the adoption of this kinematically biased
selection, leading to an incomplete picture of galaxy formation. This was also the case in the seminal paper,
\citet{els1962}, where the free-falling collapse scenario was brought by the adopted selection of halo stars based
on their high-proper motions. Similar cautions should be taken when selecting halo stars based on the paucity of metal
abundances, because some or a large fraction of the halo can be made of metal-rich stars as demonstrated by
\citet{bonaca2017} and \citet{belokurov2020}.

In this work, we explore a new selection scheme, or selection
criteria for halo stars, based on their characteristic distribution in the phase space defined by the integrals of motion,
combined with the chemical abundance information; the method is intended to be free from kinematical biases associated
with simple high-velocity cuts. The phase space, in which the sample stars are analyzed, is defined by the three integrals
of motion in an adopted axisymmetric Galactic potential: the total binding energy, $E$, the vertical angular momentum, $L_z$,
and the third integral, $I_3$. We adopt a gravitational potential of St\"ackel type, so that $I_3$ can be written in
analytical form \citep[e.g.,][]{dezeeuw1985}.

In order to determine this new set of selection criteria in ($E$, $L_z$, $I_3$), we make use of a large sample
of SDSS-SEGUE DR7 and APOGEE DR16 catalogs, matched with {\it Gaia} DR2. We then explore the most probable ranges of
($E$, $L_z$, $I_3$) for the halo system that allows a separation from rotating disk components, and derive its general
properties.

With these data sets, we also investigate the so-called \enquote{coarse-grained} phase-space distribution of
halo stars, where the averaged properties of the halo system, over the phase-space, are close to a dynamically
steady state \citep{binney2008}. The \enquote{coarse-grained} distribution differs from the \enquote{fine-grained} one,
which still contains non-relaxed small-scale substructures; the identification of these clumpy features in phase
space has been a central subject in recent literature, in particular since {\it Gaia} DR2 was released.
Here, rather than looking for such substructures, we focus on the global phase-space distribution of halo stars
and attempt to find the most likely functional form for its representation.

The paper is organised as follows. Section 2 describes the selection method for halo stars, the adopted mass model and
data. In Section 3 we analyze the distribution of stars in the phase-space and its dependence on metallicity, [Fe/H],
as well as $\alpha$-elements abundance. Section 4 present an extended discussion, including the halo duality, the
connection with the Sagittarius dwarf spheroidal galaxy (Sgr dSph) and globular clusters, and the in-situ stellar halo. 
In particular, we provide the possible relation between the Sgr dSph and the discontinuous phase-space distribution
of metal-poor halo stars. Implications for the formation of the stellar halo are also discussed in Section 5.
Section 6 presents the summary of this work and prospects.

\section{Understanding the stellar halo: selection method, mass model and adopted data}

\subsection{Background}

In many recent works, the selection of halo stars has conveniently been made based on high-velocity cuts, such as
$V_{\rm rel} = |V_{\rm star} - V_{\rm LSR}| > V_{\rm lim}$, with $V_{\rm lim}$ being 180 to 220 km~s$^{-1}$, where
$V_{\rm star}$ is the three-dimensional (3D) velocity of a star, and $V_{\rm rel}$ is its relative velocity with respect to
the Local Standard of Rest (LSR), $V_{\rm LSR}$ \citep[e.g.,][]{nissen2010, bonaca2017, haywood2018, helmi2018}.
Some studies have set a cut in a tangential motion instead of a 3D velocity, due to the limited availability of
line-of-sight
velocity information. Such a selection aims to obtain a straightforward removal of stars with disk-like kinematics,
characterized by nearly circular orbits, and thus, small velocity difference with respect to
$V_{\rm LSR}$ ($\sim$ 220 km~s$^{-1}$). 

However, this selection method is highly kinematically biased against those halo stars
possessing low $V_{\rm rel}$, which likely have low orbital eccentricities, $e$. 
Also, thanks to {\it Gaia} DR2 and other recent star catalogs from ground-based observations \citep[e.g.,][]{qiu2020},
it is now possible to analyze the kinematics of stars at larger distances from the Sun. The velocity distributions
of such remote stars can differ from those observed in the vicinity of the Sun, and the overlapping fraction of halo/disk
populations in velocity space is also a function of the Galactic position.
Therefore, the simple velocity cut of $V_{\rm rel} > V_{\rm lim}$ for the selection of halo stars is not well-founded.

The kinematics of stars are generally described in the 6D phase space, $(\bmf{x}, \bmf{v}, t)$ and, consequently, the
velocity distribution of stars depends on positions, $\bmf{x}$, and even time, $t$. Thus, the velocity-based selection
criterion that a star belongs to either disk or halo, is a function of its current position,
$\bmf{x}$, in the Galactic space.
For example, canonical thick-disk stars possessing mean azimuthal velocity of $\langle V_\phi \rangle \simeq 200$ km~s$^{-1}$
near the position of the Sun, show a finite, negative vertical gradient, $\partial \langle V_\phi \rangle/\partial z$,
such that $\langle V_\phi \rangle$ at $z = 2$~kpc is only about 160 km~s$^{-1}$, and the velocity dispersion in the $\phi$
direction, $\sigma_\phi$, also varying with $z$, is about 60 km~s$^{-1}$ at this height \citep[e.g.,][]{carollo2010}.
Other velocity dispersion components vary with the vertical distance as well. This suggests
that the simple velocity cut described above, which should be applied only to stars in the solar neighborhood, can
misclassify disk stars as halo stars, or vice versa. One may then adopt the position-dependent kinematic criteria,
but such a method depends on the adopted grids of the spatial coordinates and it is thus complex.

A useful solution is to adopt a more generalized method to characterize halo/disk stars that is less sensitive to
kinematic selections/biases. In this respect, we emphasize that an orbit-based selection in combination with other information, such as the chemical abundance, offers an ideal solution in assessing whether a star belongs to a disk or a halo component. Although the orbit-based method is dependent on the spatial distribution of the adopted sample stars, such dependence can be explicitly corrected, as demonstrated in Section 3.

\subsection{Stars in the integrals of motion space}

In general, stars in the Milky Way possess anisotropic velocity distributions and, therefore, their phase-space
distribution function depends on three isolating integrals of motion. For axisymmetric dynamical models, these integrals
are denoted as $E$, $L_z$, and $I_3$, where $E$ is the orbital energy, $L_z$ is the angular-momentum component parallel
to the $z$ axis, and $I_3$ is the third integral of motion. Thus, the orbits of stars in an axisymmetric gravitational
potentials are characterized by their distribution in a phase space defined by the integral of motion $(E, L_z, I_3)$,
where halo and disk components are expected to have their own distributions.

While $E$ and $L_z$ are classical integrals with an exact mathematical expression, the determination of the third integral,
$I_3$, has long been a central subject in Galactic dynamics: there exist no general analytical expressions for $I_3$, and
thus, its form has been investigated from numerical techniques
\citep[e.g.,][]{richstone1980,richstone1984,levison1985a,levison1985b}.
However, many axisymmetric models for the Milky Way can be approximated and described with a gravitational potential of
St\"{a}ckel form, where the Hamilton-Jacobi equation separates in ellipsoidal coordinates \citep{dezeeuw1985}.
In this case, $I_3$ can be explicitly given in analytical form and the orbit of each star is described by three integrals
of motion, $E$, $I_2 = L_z^2 / 2$, and $I_3$, which is
a generalization of $L^2 – L_z^2$ (where $L$ is the total angular momentum) in the limit of spherical symmetry
\citep{dejonghe1988}.
The analytical expression for $I_3$ allows
a fast estimation of this important quantity for a large number of sample stars, and it is particularly advantageous in the current and future big-data era.

\subsection{Selection method}

In this work, we adopt a St\"{a}ckel form for the Galactic potential and estimate $I_3$ for each star. We then investigate
the distribution of the adopted samples in the $(E, L_z, I_3)$ space, and its dependence on the metallicity, [Fe/H].
The most likely set of ranges in $(E, L_z, I_3)$ for the kinematic selection of halo and disk stars are then explored.
In particular, in addition to the $E$ vs. $L_z$ diagram commonly employed in previous studies,
we consider the distributions of stars in the $I_3$ vs. $L_z$, as well as the $I_3$ vs. $E$ diagram.

We note that in the frequently used angular-momentum diagram defined as $L_{\perp} \equiv (L_{x}^{2} + L_{y}^{2})^{1/2}$
vs. $L_{z}$, the stellar positions are changing with time because {\it $L_{\perp}$ is not an integral of motion in a
non-spherical case}, although this quantity, as well as $L_z$, can be straightforwardly determined without assuming
a gravitational potential. The orbital eccentricity, $e$, in the radial direction, $r$, and the maximum vertical distance
of an orbit from the Galactic plane, $z_{\rm max}$, are also frequently used to characterize
the orbital properties, but strictly speaking, they are not ideal parameters for the selection of disk/halo stars because
each of these depends on a combination of $E$, $L_z$, and $I_3$, and they are related to each other.
For instance, low-$e$ stars can have both large and small $z_{\rm max}$, so that by simply assigning low-$e$ stars to
a disk component leads to a contamination from those halo stars that possess low $e$ and/or large $z_{\rm max}$.
High-$e$ stars, say $e > 0.7$, are likely stellar members of a halo component, but thick-disk stars at $z \sim 2$~kpc
can have $e$ as large as 0.8 at the 2$\sigma$ significance of the $V_\phi$ distribution. Thus, the value of $e$
alone cannot be used for the classification of disk/halo stars.
\newpage

\subsection{Mass model and orbital structure}

In this subsection, we briefly describe the mass model that leads to the gravitational potential of St\"ackel type adopted
in this paper and the main properties of the orbital parameters that are relevant to this analysis. More details are given
in the Appendix~\ref{section: appendix_A}.
Although the essence of this type of a Galaxy model has already been discussed in earlier papers, its basic properties are
described below for the sake of completeness.


We adopt the axisymmetric Galactic potential of St\"ackel type, which was originally introduced by \citet{sommer1990}
and later utilized by \citet{chiba2000}.
This model consists of a disk and dark halo, where both are of St\"ackel type. The former is given by a highly flattened,
perfect oblate spheroid, resembling a Kuzmin disk, and the latter represents nearly an isothermal sphere given by a slightly
flattened, oblate spheroid, which is originally developed by \citet{dezeeuw1986}. This type of potential is
defined in spheroidal coordinates $(\lambda, \phi, \nu)$, where $\phi$ corresponds to the azimuthal angle in usual cylindrical
coordinates $(R,\phi,z)$, and $\lambda$ and $\nu$ are representing the surfaces of a spheroid and hyperboloid, respectively, in the meridional plane (See Figure~\ref{fig: coord} in the Appendix~\ref{section: appendix_A}).
The foci on the $z$ axis at $z = \pm \Delta$ fix the coordinate system and
$(\lambda, \nu)$ are bounded with $-\gamma \le \nu \le -\alpha \le \lambda$, where $\alpha$ and $\gamma$ are constants. Note that
$\lambda = {\rm const.}$ is approaching to $r = {\rm const.}$ in spherical coordinates at large $r$ and $\nu = {\rm const.}$ is
nearly $\theta = {\rm const.}$, an angle from the Galactic plane ($z=0$)\footnote{Here, for the convenience of the following
discussion, we define $\theta$ as an angle from the equatorial plane instead of a usual polar angle in spherical coordinates.},
but not exactly the same (Figure~\ref{fig: coord} in the Appendix~\ref{section: appendix_A}).

The orbits in this potential posses the three integrals of motion, $E$, $I_2 = L_z^2 / 2$, and $I_3$. The third integral,
$I_3$, is explicitly written in analytical form:
\begin{equation}
I_3 = \frac{1}{2}(L_x^2 + L_y^2) + \Delta^2 \left[ \frac{1}{2}v_z^2 - z^2 \frac{G(\lambda)-G(\nu)}{\lambda - \nu} \right] \ ,
\label{eq: I_3}
\end{equation}
where $L_x$ and $L_y$ are the angular momentum components in the $x$ and $y$ directions, respectively, and $v_z$ is
the velocity component in the $z$ direction. $G(\tau)$ with $\tau = \lambda, \nu$ is an arbitrary function representing
the gravitational potential as given in the Appendix. This expression suggests that $I_3$ becomes $\frac{1}{2}(L^2 - L_z^2)
= \frac{1}{2} L_\perp^2$ in the limit of $\Delta \to 0$, i.e., spherical symmetry. This in turn indicates that $L_\perp$
is not an integral of motion in a non-spherical potential as stated in the previous subsection: for instance, even if
$L_\perp \equiv (L_x^2 + L_y^2)^{1/2} = 0$, $I_3$ generally takes a non-zero value for a star with $v_z \ne 0$ and/or $z \ne 0$.
The boundaries of the orbit in the meridional plane are along $\lambda = {\rm const.}$ and $\nu = {\rm const.}$, i.e.,
$-\gamma \le \nu \le \nu_{+}$ and $\lambda_{-} \le \lambda \le \lambda_{+}$, and these boundaries
$(\nu_{+},\lambda_{-},\lambda_{+})$
are a function of $(E,I_2,I_3)$ \citep{dejonghe1988}. The third integral, $I_3$, is especially important for
constraining $\nu_{+}$, or nearly the maximum angle of the orbit with respect to the equatorial plane, in contrast to the
spherically symmetric case.

\begin{figure}[t!]
\centering
\includegraphics[width=80mm]{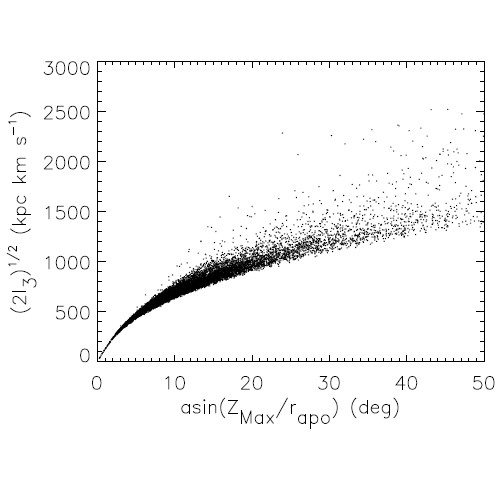}
\caption{
The relation between \I \ and the maximum orbital angle from the Galactic plane, $\theta_{\rm orb}$, estimated by
$\arcsin (z_{\rm max} / r_{\rm apo})$. The data set is the SDSS DR7 calibration stars.
It follows that \I \ correlates well with $\theta_{\rm orb}$ up to \I $\simeq 1000$ \kpckms, but
this correlation is subject to a large dispersion beyond this value, because of a finite mismatch
between the spheroidal and spherical coordinates due to the non-zero foci at $z=\pm \Delta$ in the former
(see Figure~\ref{fig: coord} in the Appendix~\ref{section: appendix_A}).
}
\label{fig: angle}
\end{figure}

To get more insights into $I_3$ in terms of more familiar orbital parameters, we plot, in Figure~\ref{fig: angle},
the values of \I \ as a function of the maximum orbital angle from the Galactic plane, $\theta_{\rm orb}$,
which may be estimated by $\arcsin (z_{\rm max} / r_{\rm apo})$, where $r_{\rm apo}$ stands for the apocentric Galactocentric
radius in spherical coordinates.
We consider $I_3$ in the form of \I \ as this is in the dimension of an angular momentum, and the data used here
are the SDSS DR7 calibration stars (see below). Inspection of Figure~\ref{fig: angle} reveals that \I \  correlates well with
the maximum orbital angle,
$\theta_{\rm orb}$, up to \I $< 1000$ \kpckms \ (corresponding to $\theta_{\rm orb} < 15^\circ-20^\circ$),
but this correlation starts to have a large dispersion beyond \I $=1000$ \kpckms, because there is a finite mismatch
between the spheroidal and spherical coordinates due to the non-zero foci at $z=\pm \Delta$ in the former
(see Figure~\ref{fig: coord} in the Appendix~\ref{section: appendix_A}).
As a guide, \I $= 500$ \kpckms \ corresponds to $\theta_{\rm orb} \simeq 5^\circ$, \I $=1000$ \kpckms \ is
$\theta_{\rm orb} \simeq 15^\circ-20^\circ$, and \I $> 1000$ \kpckms \ is $\theta_{\rm orb} > 20^\circ$.

\begin{figure*}[t!]
\centering
\includegraphics[width=170mm]{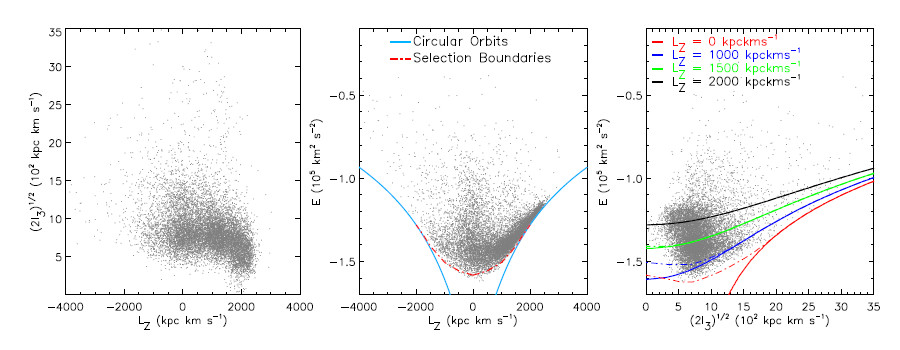}
\caption{Distributions of the SDSS DR7 calibration stars in the phase-space $(E, L_z, I_3)$ shown in terms of three diagrams,
\I \ vs. $L_z$ (left), $E$ vs. $L_z$ (middle), and $E$ vs. \I \ diagrams (right panel).
The grey dots represent the entire sample. In the middle panel, the blue solid lines denote circular orbits in the equatorial
plane of the gravitational potential. The dash-dotted line is the boundary induced by the SDSS DR7 survey's footprint
(fan shape, Fig.\ref{fig: footprints} of Appendix~\ref{section: appendix_B1}), below which the orbits are missed from this survey volume.
In the right panel, the color-coded solid lines show the boundaries of the orbits in the $E$ vs. \I \ space for a given
$L_z$, derived from Equation~(\ref{eq: momentum}), (red: $L_z = 0$, blue: 1000 \kpckms, green: 1500 \kpckms,
black: 2000 \kpckms), where $I_3 = 0$ corresponds to a circular orbit for the case of a non-zero $L_z$.
The dash-dotted lines show the boundaries associated with the sample's survey footprints
: the stellar orbits below this boundary, depicted for each $L_z$ (with the same color-coding as the solid line), are missed due to
the selection effect of the survey's footprint. The dashed-dotted line is overlapped with the solid line only
for the case of $L_z=2000$ \kpckms. Note that, the $I_3$ = 0 point for each dash-dotted line in the right panel, and for a given $L_z$, is located at the lowest $E$, represented by the red dash-dotted line in the middle panel. Thus, when $L_z$ $\lesssim$ 1500 \kpckms, the $I_3$=0 orbits are eccentric, since they are located above the blue solid line (circular orbit).
}
\label{fig: phase_selection}
\end{figure*}
\newpage


\subsection{Data: {\it Gaia} DR2, SDSS DR7 calibration stars, SDSS DR7 full sample, and SDSS DR16}

When selecting halo stars, the use of different star catalogs can sometimes lead to different results. In some cases,
this is due to the individual footprint, or sky coverage, associated to each survey. In this analysis we take into account
of this issue by employing two distinct data sets cross-matched with {\it Gaia} DR2, namely, the Sloan Digital Sky Survey
(SDSS) DR7 calibration stars and the SDSS DR16 \citep{ahumada2019}, which includes data from the Apache Point Observatory
Galaxy Evolution Experiment (APOGEE), where the latter covers the APOGEE footprints over disk regions, which are only
partially included in SDSS DR7.

We first make use of the SDSS DR7 \citet{yanny2009} calibration stars (see Appendix for description). The sample consists of $\sim$ 40,000 stars with
stellar parameters obtained by employing the SEGUE Stellar Parameter Pipeline
(SSPP; \citealt{lee2008a, lee2008b}). These parameters, as well as the $\alpha$-elements abundance are available in the
SDSS archive.\footnote{http://skyserver.sdss.org/dr7}. The average error on [Fe/H] is 0.2 dex at SNR = 10, 0.14 at
SNR = 15,  0.1 SNR = 20, 0.08 at SNR = 25, and 0.07 at SNR = 30 \citep{lee2008a}. In this sample only 14\% of the stars
has SNR $<$ 20 (with $\sigma_{\rm [Fe/H]}$ = 0.1$-$0.2),
the remaining 86\% has SNR $>$ 20 ($\sigma_{\rm [Fe/H]}$ = 0.07$-$0.08). In case of the $\alpha$-abundances the
uncertainty is is 0.06 dex at SNR = 50, and $<$ 0.1 dex at SNR = 20 \citep{lee2011a}.

The entire sample of SDSS DR7 stars is also considered for comparison, and it consists of 65,500 stars selected by
requiring, S/N $>$ 40, as well as reliable stellar parameters and $\alpha$-abundances.

Finally, we selected 70,000 unique stars from SDSS Data Release 16 by applying a series of cuts to remove stars with unreliable stellar parameters, or elements abundances. The details of these selections are reported
in the Appendix~\ref{section: appendix_B1}.

The samples are cross-matched with the Gaia DR2 database to retrieve accurate positions, trigonometric parallaxes, and proper motions, using the CDS (Centre de Données Astronomiques de Strasbourg)
X-Match service, and adopting a very small search radius (0.6''- 0.8'') to avoid duplicates. 
The match provides positions, parallaxes, and 
proper motions for all of the stars in both samples. We then select stars with relative parallax errors of $\sigma_{\pi}/\pi <$ 0.2, and derive their distance estimates using the relation $d = 1/\pi$. This selection reduces the number of stars to 10,820 for the DR7 calibration stars, and 62,060 for the APOGEE sample. In case of SDSS-SEGUE DR7 full sample the number is reduced to 46,500.
The majority of stars in this final samples have errors on proper motions below 0.2 mas yr$^{-1}$. We also adopted a parallax
zero-point offset of $\delta_{\pi}$ = $-$0.05 mas (see Appendix~\ref{section: appendix_B3} for a discussion).

Radial velocities for stars in the DR7 sample are derived from matches to an external library of high-resolution spectral templates with accurately known velocities, degraded in resolution to match the SDSS spectra (see \citealt{lee2008a}). The typical
precision is on the order of 5$-$20 km s$^{-1}$ (depending on the S/N of the spectra). In case of APOGEE stars, initial measurement of the radial velocity for each star is made by cross
correlating each spectrum with the best match in a template library, then radial velocities for each visit are derived again when the visit spectra are combined. Typical accuracy is of the order of 0.35 km s$^{-1}$ \citep{nidever2015}, however, we found that the majority of the stars in our sub-sample have an accuracy $<$ 0.2 km s$^{-1}$.

The full space and orbital motion is derived by combining the observables obtained from {\it Gaia} DR2, i.e., positions, distances, and
proper motions ($\alpha$, $\delta$, $\pi$, $\mu_{\alpha}$, $\mu_{\delta}$), with the radial velocities provided by SDSS. 
The velocities calculated in the Local Standard Rest (LSR), assumed to be rotating at 220 km s$^{-1}$,
are referred to as $(U, V, W)$ which are corrected for the motion of the Sun by adopting the values
($U, V, W$) = ($-$9,12,7) km s$^{-1}$ (Mihalas \& Binney 1981)\footnote{More recent evaluations of the LSR and solar are
available, however we adopt these values for consistency with the \citep{carollo2007,carollo2010} analyses}. The velocity
component $U$ is taken to be positive in the direction toward the Galactic anti-centre, the $V$ component
is positive in the direction toward Galactic rotation, and the $W$ component is positive toward the north Galactic pole.

\begin{figure*}[hbt!]
\vspace{-7cm}
\hspace{1.5cm}
\includegraphics[width=220mm]{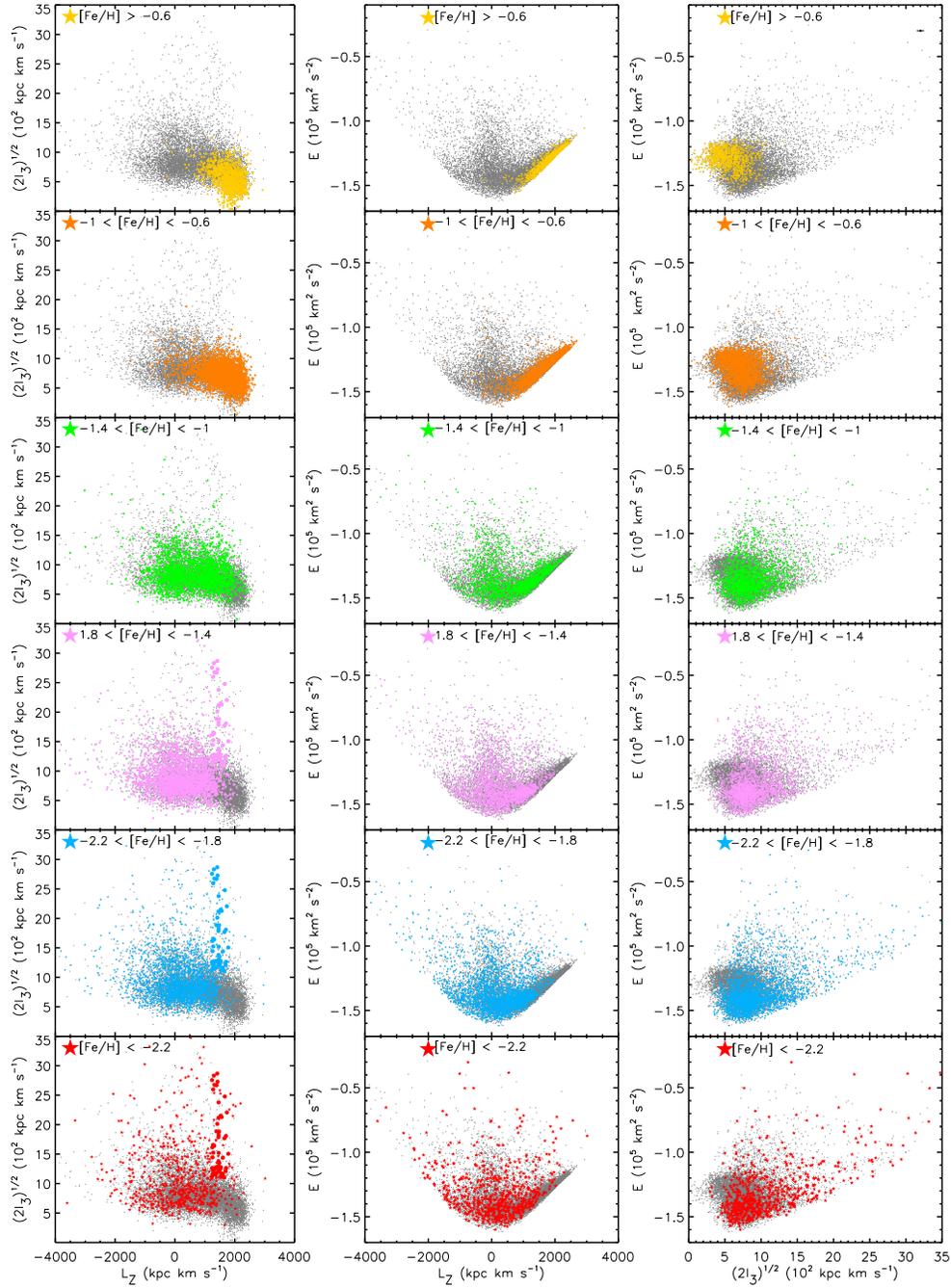}
\caption{
Distributions of the SDSS DR7 calibration stars in the phase-space $(E, L_z, I_3)$ shown in terms of three diagrams, \I \ vs. $L_z$ (left), $E$ vs. $L_z$ (middle), and $E$ vs. \I \ diagrams (right panel). In this figure,
six metallicity ranges are shown, [Fe/H]$>-0.6$ (first row of panels), $-1<$[Fe/H]$<-0.6$ (second row of panels), $-1.4<$[Fe/H]$<-1$ (third row of panels), $-1.8<$[Fe/H]$<-1.4$ (fourth row of panels), $-2.2<$[Fe/H]$<-1.8$ (fifth row of panels), [Fe/H]$<-2.2$ (sixth row of panels). In the left fourth to sixth panels, 
stars in the region where the "trail" is identified, 1200 $< L_z <$ 1800 \kpckms and $1100 <$ \I $< 3500$ \kpckms,
are shown with a filled circle symbol.  
}
\label{fig: phase1}
\end{figure*}

\section{Distributions of stars in phase-space}

\subsection{SDSS DR7 calibration stars}

\subsubsection{General properties of the phase-space distribution}

Figure~\ref{fig: phase_selection} shows the global distributions of the SDSS DR7 calibration stars in the phase space
defined by the three integrals of motion $(E, L_z, I_3)$, and represented by grey dots. In this figure, the left, middle, and
right panels show the \I \ vs. $L_z$, $E$ vs. $L_z$, and $E$ vs. \I \ diagrams, respectively.

Inspection of \I \ vs. $L_z$ diagram (left panel in Figures~\ref{fig: phase_selection}), reveals that stars with $L_z \simeq 2000$ \kpckms possess low values of the third integral of motion,
\I \ $\lesssim$ 400 \kpckms, while stars with angular momentum, $L_z < 2000$ \kpckms, have always
\I \ above $\sim$ 400 \kpckms.
In particular, there are no stars having both $L_z \simeq 0$ and \I (or $I_3$) $\simeq 0$.

As widely investigated and recognized in previous works, the distribution of stars in the $E$ vs. $L_z$ diagram
(middle panel) is bounded in a parabola shape (blue solid lines), corresponding to the circular orbits
in the equatorial plane of the gravitational potential. The dash-dotted curve in the ($E$, $L_z$) distribution represents the boundaries
of the orbits associated to the SDSS calibration stars footprint, 
and suggests that this local sample misses stars with $E \lesssim -1.6 \times 10^5$ km$^2$~s$^{-2}$,
whose apocentric radii are much below $R = R_{\odot}$.
Stars with low binding energies, $E < -1.5 \times 10^5$ km$^2$~s$^{-2}$, and low $|L_z|$
possess eccentric orbits located within the position of the Sun, and
their characteristics will be discussed in Section 4.4.2 as candidate in-situ halo stars.

In the $E$ vs. $L_z$ diagram, the elongated feature around $L_z=0$ is the Gaia-Enceladus (GE) debris \citep{helmi2018}
or Sausage structures \citep{belokurov2018}. The diagram is also characterized by a large number of highly
retrograde ($L_z < 0$), and high energy stars.

The $E$ vs. \I distribution (right panel of Figure~\ref{fig: phase_selection}), exhibit a parabola-shape boundary as well. In this panel, the color-coded solid lines show the theoretically allowed boundaries for
all the orbits at a given $L_z$, which are solutions of $p_\tau^2 = 0$ with $\tau = \lambda, \ \nu$ in
Equation~(\ref{eq: momentum}) (Appendix \ref{section: appendix_A}),
and $I_3 = 0$ corresponds to a circular orbit for the case of a non-zero $L_z$.

It is important to note that, in contrast to the $E$ vs. $L_z$ diagram,
where the bottom of the parabola exhibits a value of $L_z = 0$, the bottom of the parabola in the $E$ vs. \I distribution (the lowest $E$), is located
at a non-zero value of \I, of the order of $\sim 700$ \kpckms.
In fact, stars with the lowest range of energy, $E < -1.5 \times 10^5$ km$^2$~s$^{-2}$,
tends to populate the interval $400 \lesssim$ \I $\lesssim 1200$ \kpckms.
These values of \I \ correspond to orbital angles in the range of
$5~{\rm deg} \lesssim \theta_{\rm orb} \lesssim 20-30~{\rm deg}$, and $\theta_{\rm orb} \simeq 10$~deg
for the lowest $E$ (See Figure~\ref{fig: angle}).


One possible reason for the non-zero \I \ at the lowest $E$ (and $L_z \simeq 0$), may be due to the fan shape of
the SDSS footprints, which lack of stars at small radii, $R$, ($R < R_{\odot}$), and low $z$
(See Figure~\ref{fig: footprints} in the Appendix~\ref{section: appendix_B1}). This implies that stars with
small orbital angle, $\theta_{\rm orb}$, at small $R$, are outside the sampling volume of the DR7 survey.


To test a possible selection effect on such low \I stars, we derive the boundaries of the orbits
in the $E$ vs. \I  diagram associated with the sample's survey footprints.
Such boundaries are expressed by the following approximate formula:
$|z|$ {\rm (kpc)} $\ge -$ 0.980 $R$ {\rm (kpc)} + 7.882, and $R >$ 6.8~kpc inside the solar radius
(Figure~\ref{fig: footprints} in the Appendix~\ref{section: appendix_B1}).
The former equation corresponds to the line, which nearly traces the lowest $|z|$ coordinates of the sample
inside the solar position so as to pass $(R,z) \simeq (8, 0)$~kpc and $(6, 2)$~kpc.

The derived boundaries are shown by the dash-dotted lines for each of a given $L_z$ in the $E$ vs. \I \ diagram,
with the same color-coding as the solid lines: the stellar orbits below the boundary are missed due to the selection
effect of the survey's footprint in the SDSS DR7 sample. 
It is worth noting that the case of $L_z = 0$ (red dash-dotted line) reproduces very well
the boundary for stars with \I $> 700$ \kpckms.
It is also clear that while the stars with $E < -1.7 \times 10^5$ km$^2$~s$^{-2}$ are missed by the survey footprint selection, the low $I_3$ stars (\I $< 500$ \kpckms) can be definitely detected, if they exist, within the survey volume, because such low $I_3$ stars have highly elongated, eccentric orbits near the Galactic plane, so that
their apocentric radii can reach the current local survey volume, if they really exist. 
This demonstrates that the lack of low $I_3$ stars at $E < -1.4 \times 10^5$ km$^2$~$^{-2}$ in the $E$ vs. \I diagram {\it is real}, and definitely not due to the selection effect induced by the survey's footprint, and we will show
in the next subsection that the absence of low $I_3$ stars with high orbital eccentricities is an essential property
of metal-poor halo stars. We however note that the current survey volume misses low $I_3$ stars with low orbital
eccentricities, such as those with nearly circular orbits and possesing low orbital angles inside the solar position. The
possible presence of such stars, and low metal abundances, will be discussed in Section 4.4.3.

To get further insights into the surveys selection effect, we also consider the DR16 (APOGEE) catalog, whose
footprint includes smaller $R$ and lower $z$ (Figure~\ref{fig: footprints} in Appendix~\ref{section: appendix_B3}).
In  Appendix~\ref{section: appendix_B4}, Figure~\ref{fig: DR16_phase1} and \ref{fig: DR16_phase2}, show the phase-space
distribution of APOGEE stars. The right panels of Figure~\ref{fig: DR16_phase1} ($E$ vs. \I diagram) show that stars with
[Fe/H] $> -$0.6 have \I = 0 \kpckms in the range of energy,
$-1.5 \times 10^5$ km$^2$~s$^{-2}$ $< E$ $< -1.1 \times 10^5$ km$^2$~s$^{-2}$.
These are disk stars with nearly circular orbit in the Galactic plane, and orbiting at small $R$.
Such stars are not present in the SDSS DR7 footprint, as can be inferred by examining Figure~\ref{fig: phase_selection}. 
On the other hand, metal-poor stars appear to have a non-zero \I value at the lowest $E$, although the paucity of halo
samples in the APOGEE survey prevents from investigating in more details this property.
The fact that \I exhibits a non-zero value for metal-poor halo stars in both SDSS DR7 and Apogee DR16 surveys is a real physical property of these stars, and it is not due to a selection effect.

\subsubsection{Dependence on metallicity}


Figure~\ref{fig: phase1} shows the ($L_z$, $E$, \I) phase-space distribution of the SDSS DR7 calibration stars for the six metallicity
intervals: [Fe/H]$>-0.6$, $-1<$[Fe/H]$<-0.6$, $-1.4<$[Fe/H]$<-1$, $-1.8<$[Fe/H]$<-1.4$, $-2.2<$[Fe/H]$<-1.8$, and
[Fe/H]$<-2.2$. The grey dots show the entire sample, while the color-coded symbols represent sub-samples in the various
ranges of metallicity, as indicated in the legends of each panel.

The uncertainties on
the derived orbital parameters due to the observational errors
have been estimated through a Monte Carlo simulation (100
realizations for each star), and they are listed in Table 1.

In the highest metallicity range ([Fe/H]$>-0.6$, top panels of Figure~\ref{fig: phase1}, dark yellow symbols) many stars
have large rotational velocity (high values of $L_z$), and possess values of \I \ below 500 \kpckms, but all of them have
\I \ below 1000 \kpckms, and $E$ below $-1.2 \times 10^5$ \km. These stars, are distributed along a parabola with nearly
circular orbits in the $E$ vs. $L_z$ diagram, and are dominated by stellar members of the thin- and thick-disk components.
As the metallicity decreases to $-1<$[Fe/H]$<-0.6$ (2nd row in Figure~\ref{fig: phase1}, orange symbols), the number
of stars having \I \ around 1000 \kpckms\ , and beyond, increases, as well as those with lower $L_z$, including some stars
with $L_z$ as small as 0. In this range of metallicity, the overlapping thick disk and metal-weak thick disk (MWTD;
\citealt{carollo2019}), which are predominantly in circular orbits, dominate the distribution, with some halo stars
contamination.

In the two intermediate ranges of metallicity represented by $-1.4<$[Fe/H]$<-1$ (3rd row in Figure~\ref{fig: phase1},
green symbols) and $-1.8<$[Fe/H]$<-1.4$ (4th row in Figure~\ref{fig: phase1}, pink symbols), stars exhibit progressively
lower $L_z$, and there are almost no stars with \I $<$ 500 \kpckms. In these metallicity intervals, many stars have
\I $>$ 1000 \kpckms, reaching values up to $\sim$ 3000 \kpckms, in contrast to the higher metallicity
ranges ([Fe/H] $>-0.6$) where almost no stars possess \I \ $>$ 1000 \kpckms. Also, while the number of stars with
large $L_z$ and low \I \ decreases, many stars with lower $L_z$ tend to populate the elongated
feature in the center of the $E$ vs. $L_z$ diagram, over the range of $-500 < L_z < 500$ \kpckms, and possessing an
extended distribution in both $E$ and \I. This feature is dominated by the GE debris stars \citep{helmi2018}, or Sausage
structure \citep{belokurov2018}. Note that the range of metallicity, $-1.8<$[Fe/H]$<-1.4$, matches with that of the
{\it inner halo} stellar population discussed in \citet{carollo2007, carollo2010}, whose metallicity peak is
[Fe/H]$\sim -1.6$. In these intermediate metallicity intervals, there are also many stars with retrograde motion and
high energy.

It is interesting to notice that in the \I \ vs. $L_z$ distribution, and in the metallicity range of $-1.4<$[Fe/H]$<-1$
(3rd row and left panel in Figure~\ref{fig: phase1}, green symbols), at $700 < L_z < 1500$ \kpckms,
there exists a distinct distribution of stars, which contains both the MWTD and halo stars.
The MWTD has a peak of metallicity of [Fe/H] $\sim -$1.0 and $\langle L_z \rangle \sim 1200$ \kpckms \ \citep{carollo2019};
this feature becomes slightly weaker as the metallicity decreases to $-1.8<$[Fe/H]$<-1.4$.

Over the range of $-2.2<$[Fe/H]$<-1.4$, another notable feature is clearly present in the \I \ vs. $L_z$ diagram:
the elongated distribution of stars from $(L_z, (2I_3)^{1/2}) \simeq (1800, 400)$ \kpckms to $(1000, 3500)$ \kpckms.
Such a feature was originally identified and called a 'trail' feature in the $L_\perp$ vs. $L_z$ diagram by
\citet{chiba2000} (see their Figure 15). We will discuss the origin of this structure in Section 4.

At lower metallicity, $-2.2<$[Fe/H]$<-1.8$ and [Fe/H]$<-2.2$ (5th and 6th rows in Figure~~\ref{fig: phase1},
blue and red colors, respectively), the majority of the stars are characterized by $L_z < 1500$ \kpckms, and
\I $> 500$ \kpckms. We note in particular the rather sharp boundary at $L_z \simeq 1500$ \kpckms, and
\I $\simeq 500$ \kpckms, in the distributions of these low metallicity stars, which are dominated by halo stars.
By comparing with Figure~\ref{fig: phase_selection}, it is clear that the discontinuity in the distribution of the metal-poor
stars at $L_z \simeq 1500$ \kpckms \ is an intrinsic property of the metal-poor population, and it is not due to any selection effect of
the survey volume. The distribution below $L_z = 1500$ \kpckms is extended over large ranges of $E$ and \I,
and there are stars with retrograde motion and high energy, reaching up to $E = -0.4 \times 10^5$ km$^s$~s$^{-2}$,
as quantified below (Figure~\ref{fig: histo_phase}).

\begin{deluxetable}{ccccccc}
\tablecolumns{7}
\tablewidth{0pt}
\tablecaption{Average errors on orbital parameters}
\tablehead{
\colhead{$\sigma_{\rm I_3}/I_3$} &
\colhead{$\sigma_{\rm (2I_3)^1/2}$} &
\colhead{$\sigma_{\rm E}$} &
\colhead{$\sigma_{L_{\rm Z}}$} &
\colhead{$\sigma_{r_{\rm apo}}$} &
\colhead{$\sigma_{r_{\rm peri}}$} &
\colhead{$\sigma_{ecc}$}\\
\colhead{} &
\colhead{(\kpckms)} &
\colhead{(10$^{5} $km$^{2}$~s$^{-2}$)} &
\colhead{(kpc)} &
\colhead{(kpc)} &
\colhead{(kpc)} &
\colhead{} 
}
\startdata
    0.13    & 50  &  0.01  &  100  & 0.26 & 0.35 & 0.04 \\
     &   &    &    &  &  &  \\
\hline
\enddata
\end{deluxetable}

It is worth noting that stars possessing retrograde motion were already identified and described as members of
'the outer halo' component in the original sample employed by \citet{carollo2007, carollo2010}. The retrograde motion
detected in that sample is confirmed when adopting the more precise {\it Gaia} DR2 parameters. 


\begin{figure*}[hbt!]
\centering
\vspace*{-0.4cm}
\includegraphics[width=135mm]{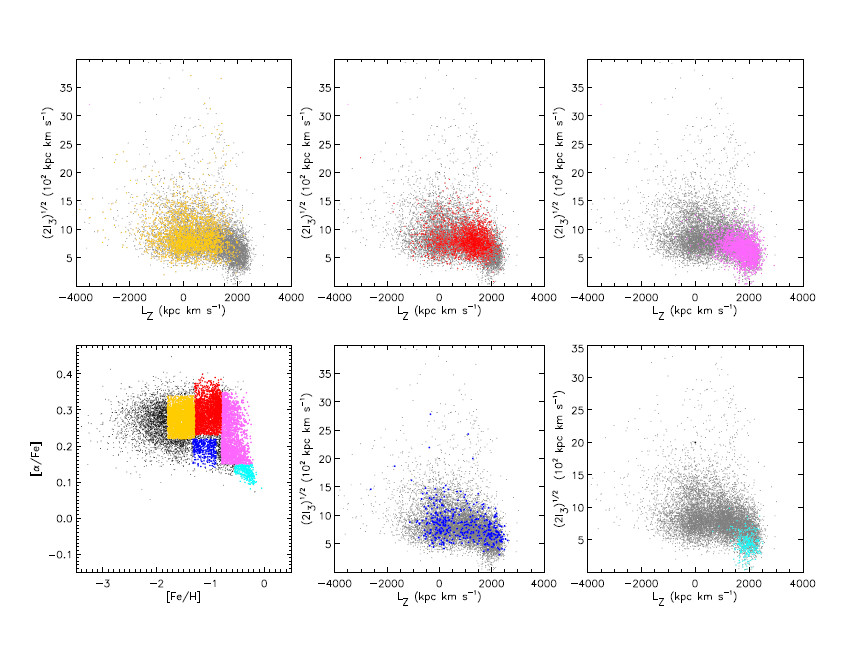}
\caption{
Distribution of DR7 calibration stars subsamples in the \I \ vs. $L_z$ diagram, selected within
color-coded regions of the [$\alpha$/Fe] vs. [Fe/H] diagram (bottom-left panel). Stars in these regions represent the thin disk (cyan), the thick disk (pink), the MWTD (red), the GE debris (yellow), and the transition
between the GE debris and the metal-poor thin disk (dark blue).
}
\label{fig: phase1_sub}
\end{figure*}
\begin{figure*}[hbt!]
\centering
\vspace*{-0.4cm}
\includegraphics[width=135mm]{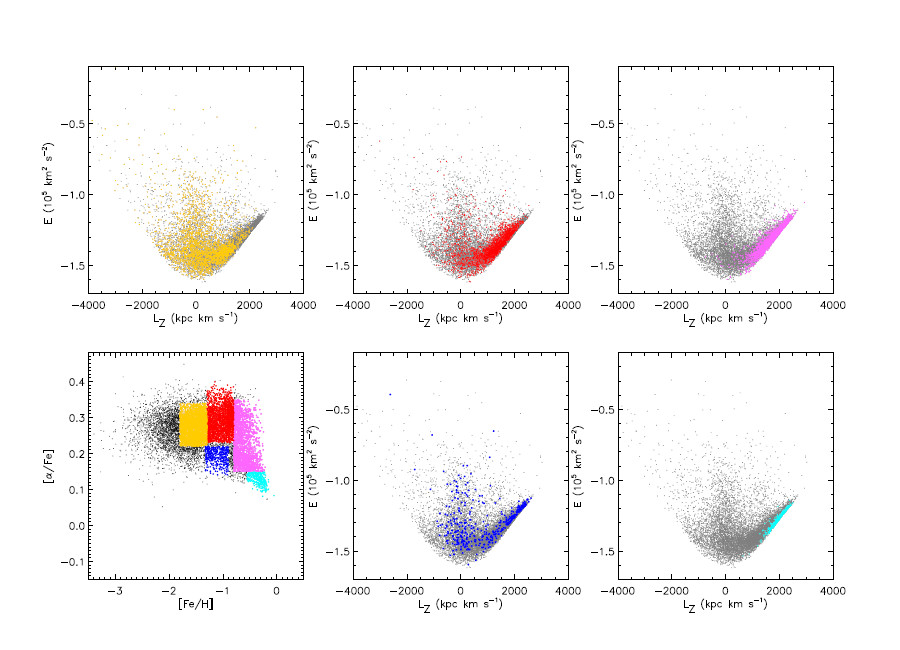}
\caption{
The same as Figure~\ref{fig: phase1_sub} but in the $E$ vs. $L_z$ diagram  for the DR7 calibration stars.
}
\label{fig: phase2_sub}
\end{figure*}
\begin{figure*}[hbt!]
\centering
\hspace{2cm}
\includegraphics[scale=0.6,angle=90]{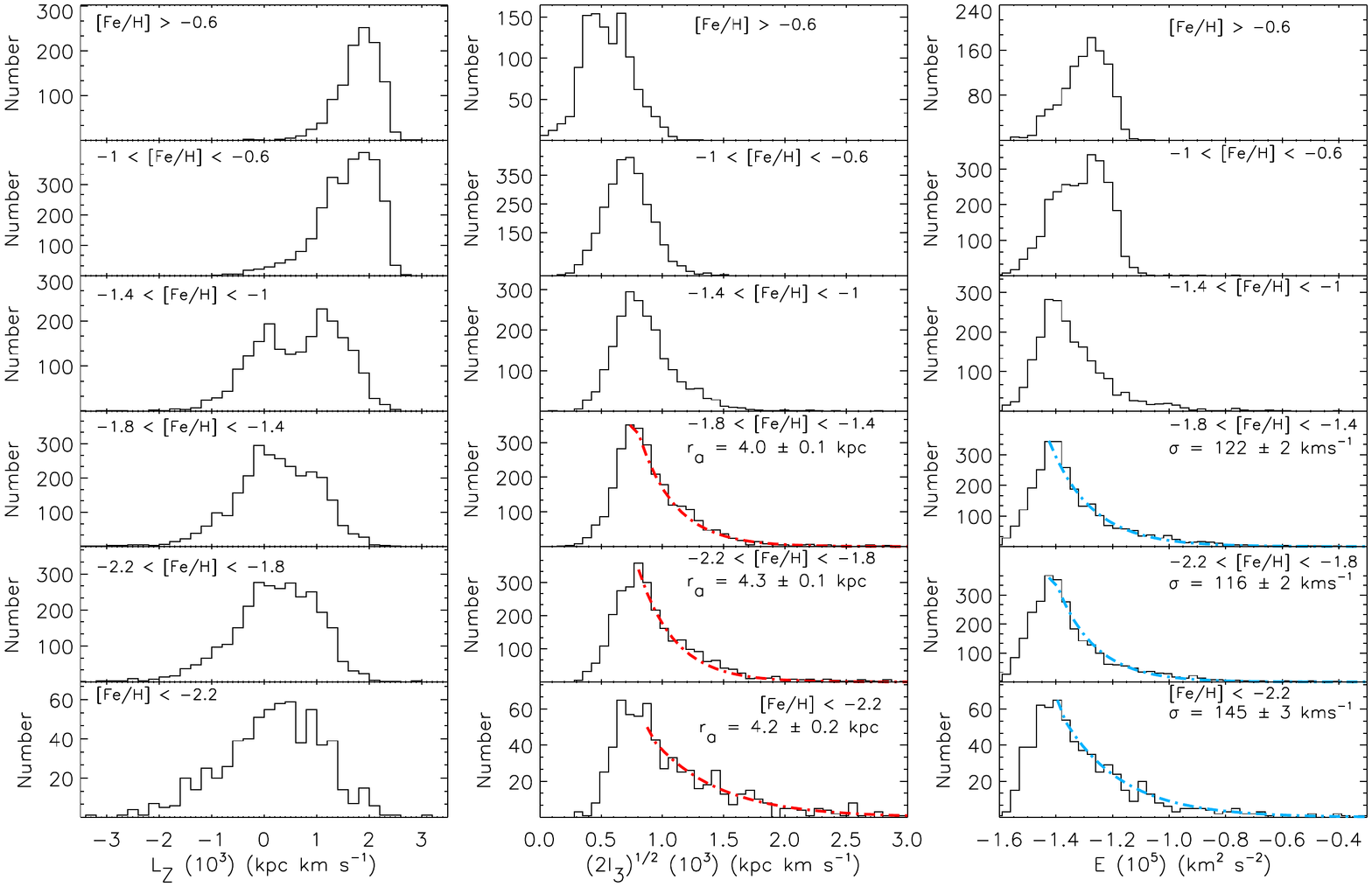}
\caption{$L_z$ (left panels), \I \ (middle panels), and $E$ (right panels) histograms in the six metallicity ranges,
for the DR7 calibration stars. The adopted bins are, 0.2 (10$^{3}$ \kpckms), 0.07 (10$^{3}$ \kpckms), and
0.03 (10$^5$ km$^{2}$ s$^{-2}$), for $L_z$, \I, and $E$, respectively.
The dash-dotted lines in the middle and right panels, and for the metal-poor ranges of [Fe/H] $<-1.4$ (4th, 5th, and 6th rows of
panels), represent the fitting to Equation~(\ref{eq: michie}) at \I $>850$ \kpckms, and $E > -1.4 \times 10^5$ km$^2$~s$^{-2}$,
respectively. The parameters $\sigma$ (1d velocity dispersion) and r$_{\rm a}$ (anisotropic radius), provided by the exponential fits, are shown in each diagram.
}
\label{fig: histo_phase}
\end{figure*}

\subsubsection{Comparison with the [$\alpha$/Fe] vs. [Fe/H] diagram}

To further investigate the phase-space distribution of the SDSS DR7 calibration stars, we select stars in specific regions of [$\alpha$/Fe] vs. [Fe/H] diagram, each of which represent likely thin disk, 
thick-disk, MWTD and GE stars, and analyze their locations in the phase-space diagram. In this exercise we use only
the \I \ vs. $L_z$ and $E$ vs. $L_z$ diagrams, as the $E$ vs. \I \ diagram has less evident features.

In Figure~\ref{fig: phase1_sub} and Figure~\ref{fig: phase2_sub}, the bottom-left panel shows the [$\alpha$/Fe] vs. [Fe/H]
diagram, where cyan and pink dots represent candidate thin disk and thick disk stars. Red and yellow dots denote the MWTD and
GE, respectively, while the dark blue dots show the transition region between the metal-poor thin disk and GE. The rest of the panels
in Figure~\ref{fig: phase1_sub} and Figure~\ref{fig: phase2_sub} show the corresponding distributions of the color-coded stars
in the \I \ vs. $L_z$, and $E$ vs. $L_z$ diagrams, respectively.

Stars with cyan symbol have large $L_z$ and low values of \I \ and they are members of the thin disk stellar population.
The color-coded pink and red stars pick up
well the thick-disk and MWTD components, and possess larger \I \  than thin-disk stars. Comparison between the top right and middle
panels reveals that the MWTD includes more stars with large values of \I than the thick disk, which suggests that stars in the MWTD component have larger orbital angles, $\theta_{orb}$ , than those in the thick disk. 
MWTD stars exhibit, on average, lower $L_z$ than thick  disk stars. The MWTD has indeed a lower mean rotational velocity, $\langle V_{\phi} \rangle \simeq 150$ \kms,
than the thick disk, $\langle V_{\phi} \rangle \simeq 180$ \kms, as described in \citet{carollo2019}.

It is interesting to notice that stars with dark blue symbols are located in the same metallicity range as those with 
red symbols ($-1.3<$[Fe/H]$<-0.9$), but populate different intervals of [$\alpha$/Fe], (dark blue, $+0.14<$[$\alpha$/Fe]$<+0.22$,
and red, $+0.23<$[$\alpha$/Fe]$<+0.4$). The middle-bottom panels of Figure~\ref{fig: phase1_sub} and
Figure~\ref{fig: phase2_sub} show that stars color-coded with dark blue symbols separate in two different distributions,
one having large $L_z$ (even larger than the cyan symbols) and low \I, and one having $L_z$ around 0 and extending towards
large values of $E$ and \I. Thus, in the region of the [$\alpha$/Fe] vs. [Fe/H] diagram color-coded with dark-blue symbols, there exist both rapidly-rotating, metal-poor stars, with values of \I and $L_z$ comparable to those of the think disk stellar
population, and likely GE debris stars.
The rapidly-rotating disk-like stars may have migrated from larger radii in the disk (more metal poor) to the solar neighborhood, where they acquired larger rotational velocities due to the angular momentum conservation, and inducing a velocity-metallicity gradient, 
$\Delta \langle V_\phi \rangle / \Delta {\rm [Fe/H]} < 0$, as observed for the thin disk sequence 
\citep{lee2011b,han2020}.
The dark-blue area in the abundance plane has been discussed in other recent works \citep[see for example,][]{hayes2018,
das2020}, suggesting the dominance of a non-rotating, accreted population of stars associated with the GE debris structure.

In the [$\alpha$/Fe] vs. [Fe/H] diagram, GE debris stars are mainly represented by the area color-coded with yellow symbols. These stars are spread around $L_z \sim 0$ and over an extended range of $E$ and \I. 
It is also important to note that some of the stars selected in the yellow area are widely distributed in both positive and
negative $L_z$ ranges, including those along the parabola shape in the $E$ vs. $L_z$ diagram.
This means that the yellow area in the [$\alpha$/Fe] vs. [Fe/H] diagram, comprises not only the GE debris
but also other parts of the halo, where stars with both prograde and retrograde rotation coexist.
\newpage

\vspace{1cm}
\subsubsection{\enquote{Coarse-grained} phase-space distribution of halo stars}

The halo system contains several sub-structures in the form of stellar streams/over-densities recognizable in
spatial distributions, or in the form of long-lived sub-structures identified in phase space. The former are relics of
relatively recent accretion/merging events of stellar systems and the latter are associated with such events
but at much earlier epochs \citep[e.g.,][]{helmi1999, freeman2002, feltzing2013}.
The halo might be entirely made of substructures \citep{naidu2020}, and it may still 
on its way of a dynamically relaxed state. However, while the halo system continues to be in the course of relaxation in a
\enquote{fine-grained} snap-shot of the phase space, such time varying features will be somehow smoothed out and closer to a steady
state in the \enquote{coarse-grained} phase space \citep{binney2008}.
Here, we attempt to quantify such a \enquote{coarse-grained}, general distribution of halo stars in the ($E$, $L_z$, $I_3$) space
by neglecting several fine sub-structures in it.

For this purpose, we show, in Figure~\ref{fig: histo_phase}, the distributions
of $L_z$ (left), \I \ (middle) and $E$ (right panels) for the six ranges of metallicity defined earlier in the paper.

At metallicity [Fe/H]$>-1$, the $L_z$ distribution (left panels) is dominated by disk-like
kinematics components (thin- and thick-disk), while at lower metallicity, $-1.4<$[Fe/H]$<-1$, the distribution splits in two components,
one with peak at $L_z \sim 1200$ \kpckms \, and dominated by the MWTD, and one with peak at $L_z \sim 0$, which represents mainly GE debris stars. As the metallicity decreases ($-1.8<$[Fe/H]$<-1.4$, fourth panel), the $L_z$ distribution is still dominated by
the GE debris, while the MWTD peak becomes weaker, and stars with retrograde motion start to appear. 
Thus, the average properties of the halo in these two last metallicity ranges are basically governed by two main structures, the MWTD and the GE debris stream.
In the lower metallicity intervals, $-2.2<$[Fe/H]$<-1.8$ and [Fe/H]$<-2.2$, the $L_z$ distribution shows
a large fraction of stars with $L_z \sim 0$ or slightly prograde, and a significant fraction of stars with highly retrograde motion.
In these metal-poor ranges, the distribution of halo stars may be simply approximated such as $f(L_z) \propto \exp(-L_z^2)$,
where the positive $L_z$ side shows a rapid decrease, truncated at $L_z \simeq 1500$ \kpckms, while the negative
$L_z$ side shows an extended tail originated by the presence of a significant number of stars with large retrograde motions.

In the \I \ and $E$ distributions (middle and right panels), the disk-like components are visible in the two top-panels of higher metallicity intervals. At [Fe/H]$<-1$ (from the third panel), the MWTD feature is absorbed by
the general halo distributions, which exhibit an exponential decrease with increasing \I \ and $E$,
starting from \I $\sim$ 800 \kpckms, and $E \sim -1.5$ - $-1.3 \times 10^5$ \km, respectively.
In fact, by considering $I_3 = L^2 / 2 - L_z^2 /2$ in the spherical limit, the function $f(E,I_3)$ may be approximated by the
so-called Michie-Bodenheimer type model \citep[see e.g.,][]{richstone1984, sommer1987, binney2008}, 

\begin{eqnarray}
    f(E,I_3) &\simeq& f(E) f(I_3)   \nonumber \\
             &\propto& \left[ \exp( - E / \sigma^2 ) - 1 \right] \exp( - I_3 / r_a^2 \sigma^2 ) 
\label{eq: michie}
\end{eqnarray}

for $E \le 0$, where $\sigma$ is a 1d velocity dispersion and $r_a$ stands for the anisotropic radius beyond which
the velocity dispersion is anisotropic\footnote{In the spherical limit, the Michie-Bodenheimer type model corresponds
to $f(E,L) \propto \left[ \exp( - E / \sigma^2 ) - 1 \right] \exp( - L^2 / 2 r_a^2 \sigma^2 )$ for $E \le 0$.}.

The analysis of the global distribution function of metal-poor halo stars, requires also to consider possible selection effect of the survey footprints in the SDSS catalog. In Sect. 3.1.1 we showed that the SDSS DR7 stellar sample lacks stars with binding energy below
$E/(10^5 {\rm km}^2~{s}^{-2}) = -1.4 \sim -1.5$, because their apocentric radii are much lower than R$_{\odot}$. This implies that the energy distribution at $E \lesssim -1.4 \times 10^5$ km$^2$~s$^{-2}$ is significantly reduced. Different is the case for the \I distribution. Indeed, the SDSS DR7 survey does not miss stellar orbits with low $I_3$, specifically, \I $ \lesssim 700$ \kpckms. Therefore, the rapid decrease of the \I distribution for metal-poor stars at \I $\lesssim 700$ \kpckms is real, and not caused by a selection effect. To take into account these properties of the $E$ and \I distribution functions, the exponential regression fit to the Equation~(\ref{eq: michie}) for the metal-poor subsamples ([Fe/H]$<-1.4$), was performed by selecting stars with \I $> 700-800$ \kpckms, and $E > -1.4 \times 10^5$ \km, 
(fourth- and fifth- middle panels of Figure~\ref{fig: histo_phase}).


The exponential regression model provides,  
$(\sigma, r_a) = (122~{\rm km~s}^{-1}, 4.0~{\rm kpc})$ and $(116~{\rm km~s}^{-1}, 4.3~{\rm kpc})$, and reproduces very well the dependence of the function $f(E,I_3)$ on $E$ and $I_3$. 
In the most metal-poor range, [Fe/H] $< -2.2$, the $E$ and $I_3$ distributions exhibit a large extension toward high values (middle and right bottom panels), thus, the model requires a systematically larger velocity dispersion to obtain a good exponential fit. We found, $\sigma$ = 145~km~s$^{-1}$, while the anisotropic radius remains the same, $r_a$ = 4.2~kpc. In Figure 6, the three bottom panels in the middle- and right-column, show the exponential curve obtained by applying the non-linear regression overlapped to the actual data (red for \I and blue for $E$).
Goodness of fit diagnostics performed for each of the six exponential fits show that the correlation between the values predicted by the model, $\exp( - E / \sigma^2 ) - 1$ for the energy, and $\exp( - I_3 / r_a^2 \sigma^2 )$ for $I_3$, and the actual data is 0.98-0.99. The normality of residuals was evaluated using the Shapiro-Wilk test \citep{ritz2008} providing that, in each case, the null hypothesis of normal distribution could not be rejected.

\begin{figure*}[hbt!]
\centering
\vspace*{-0.4cm}
\includegraphics[width=150mm]{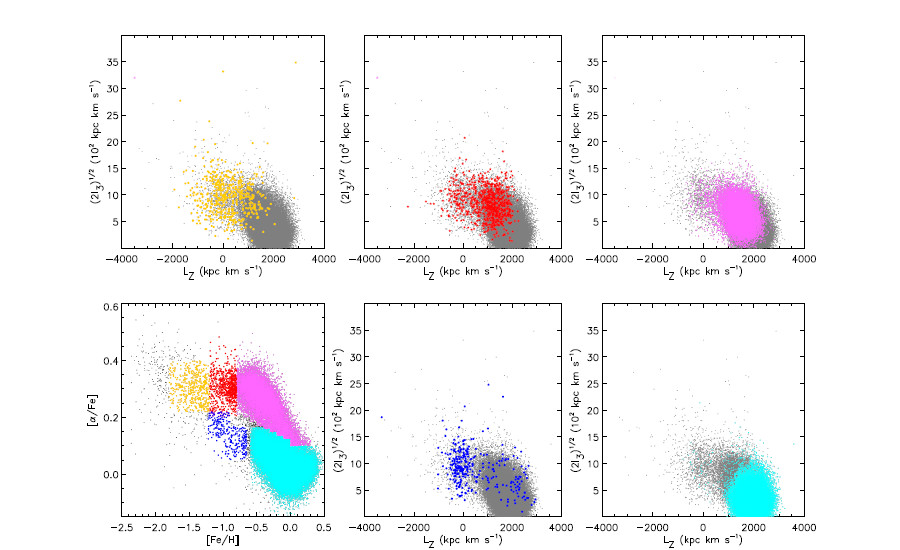}
\caption{
Distribution of the SDSS DR16 APOGEE subsamples in the \I \ vs. $L_z$ diagram, selected within color-coded regions of the [$\alpha$/Fe] vs. [Fe/H] diagram (bottom-left panel). The color coding is
the same as in Figure~\ref{fig: phase1_sub}.
}
\label{fig: DR16_phase1_sub}
\end{figure*}
\begin{figure*}[hbt!]
\centering
\vspace*{-0.4cm}
\includegraphics[width=135mm]{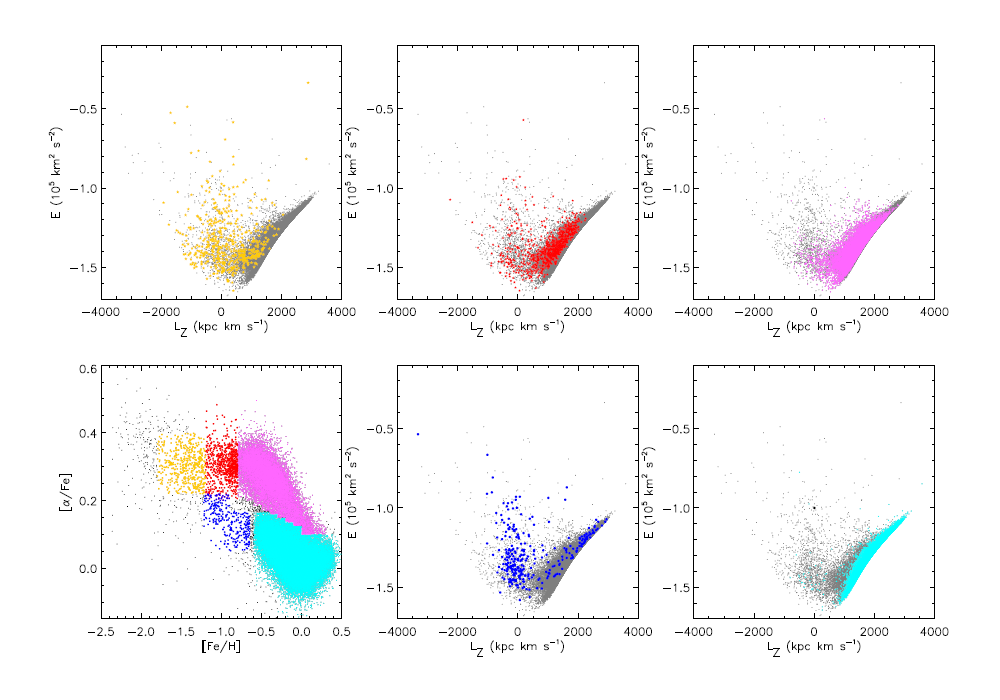}
\caption{
The same as Figure~\ref{fig: DR16_phase1_sub} but in the $E$ vs. $L_z$ diagram for the SDSS DR16 APOGEE stars.
}
\label{fig: DR16_phase2_sub}
\end{figure*}

\subsubsection{SDSS DR7 full sample}
The entire SDSS-SEGUE DR7 data set is also adopted for comparison with the SDSS-SEGUE DR7 calibration stars sample. After the application of the selection criteria, the full sample contains a larger number of stars with respect the calibration stars sample (N$_{\rm full}$ = 46,500), however, such increased number didn't add much more information on the properties of the stellar halo, already obtained with the SDSS DR7 calibration stars. The similarity of the distributions in $E-L_z$ and \I-$L_z$ as a function of the metallicity for the two samples can be assessed by comparing Figure~\ref{fig: phase1_sub} and \ref{fig: phase2_sub} of the main section with Figure~\ref{fig: DR7full_phase1_sub} and \ref{fig: DR7full_phase2_sub} in the Appendix. In case of the full data set, the sub-samples obtained for the various cuts of metallicity, contain a larger number of stars. This is valid, in particular, for the thin- and thick-disk stellar populations (cyan and pink, respectively). Nonetheless, the main features and properties of the recognized stellar populations and debris-streams, and described in section 3, remain the same.

\subsection{SDSS DR16}

Figure \ref{fig: DR16_phase1_sub} and Figure \ref{fig: DR16_phase2_sub} show the distributions of SDSS DR16 stars
in the \I \ vs. $L_z$ and $E$ vs. $L_z$ diagrams, respectively. The grey dots show the entire sample.
As for the DR7 calibration star sample, various components and features are selected by considering fiducial values
for their metallicity and $\alpha$-elements abundance, shown with color-coded dot symbols in the bottom-left panel, in a similar fashion as in Figure~\ref{fig: phase1_sub} and \ref{fig: phase2_sub}.
Cyan and pink dots represent likely thin disk and thick disk stars located in the most metal-rich range.
Red and yellow dots denote the MWTD and GE, respectively, and the blue dots show the transition region between
the metal-poor thin disk and GE. 

The DR16 sample shows a broad distribution of stars located at $L_z \sim 2000$ \kpckms, and \I $\sim 0$ \kpckms \ (Figure
\ref{fig: DR16_phase1_sub}). This is caused by the APOGEE footprint (Figure~\ref{fig: footprints}
in the Appendix~\ref{section: appendix_B3}),
which includes extended regions along the Galactic plane, and covers large ranges of the Galactocentric
radius, $3 \lesssim R \lesssim 14$ kpc. This wide footprint includes stars with large values of energy and
vertical angular momentum for circular orbits up to $E \simeq -1.0 \times 10^5$ km$^2$~s$^{-2}$, at $L_z \simeq 3000$
\kpckms \ (see Figure \ref{fig: DR16_phase2_sub}, bottom-right panel). We also notice that
at $L_z \simeq 0$ \kpckms, APOGEE stars possess slightly lower energy values than SDSS DR7 calibration stars
($E < -1.65 \times 10^5$ km$^2$~s$^{-2}$: see Figure \ref{fig: DR16_phase2_sub}). Also, the values of \I \ at
$L_z \simeq 0$ (and lowest $E$) are somewhat smaller than those for the SDSS DR7 calibration stars, suggesting lower
orbital angles, $\theta_{\rm orb}$ (compare Figure~\ref{fig: phase1_sub} and Figure~\ref{fig: DR16_phase1_sub}).
This difference is caused by the APOGEE footprint in which more gravitationally bound stars, located at smaller $R$ and $\theta_{\rm orb}$, are sampled (Figure \ref{fig: footprints}, Appendix~\ref{section: appendix_B3}).\\
While the APOGEE survey samples very well the thin- and thick-disk stellar populations, the more metal-poor halo stars are
under-represented, in particular below [Fe/H]$\sim -2$. On the contrary, in the SDSS DR7 calibration stars sample the
thin disk is under-represented, whereas the halo system is very well sampled. We thus focus on more metal-rich stars
covering the disk and disk/halo overlapping regions, which are highlighted with cyan and pink color-coded symbols.

Inspection of the bottom-right panels reveals that thin-disk stars possess large $L_z$ and small values of
\I, while  thick-disk-like stars (top-right panels) show larger \I \ and smaller $L_z$. This suggests,
as expected, that the thick disk possesses lower mean rotational velocity and larger orbital angles from the Galactic plane, than the thin disk. The MWTD stars (red symbols, top-middle panel) exhibit large values of 
\I $\geq 500$ \kpckms (average value \I $\sim 850$ \kpckms), which implies that MWTD stars possess orbits that form angles with the Galactic plane of
$\theta_{\rm orb} > 7$~deg (average value $\theta_{\rm orb} \sim 10$~deg), and systematically larger than those possessed by the thick disk (average value $\theta_{\rm orb} \sim 7$~deg). The peak of the $L_z$
distribution for the MWTD is lower than that of the thick disk, showing that the MWTD lags behind the thick disk,
in agreement with \citet{carollo2019}. Some stars selected as MWTD members have $L_z \sim 0$ \kpckms \ and
$E \sim -1.6 \times 10^5$ km$^2$~s$^{-2}$, which may be halo-star contaminants.

The bottom-middle panels in Figure~\ref{fig: DR16_phase1_sub} and \ref{fig: DR16_phase2_sub} show the transition region
between the thin disk and GE in the [$\alpha$/Fe] vs [Fe/H] diagram. Some stars fall in the areas dominated by the disk(s) populations,
while most of the stars are characterized by $L_z \sim 0$ \kpckms, and extended values of $E$ and \I, matching
the distribution of stars likely members of the GE debris (yellow dots in the top-left panels).
Note that the distribution of stars selected as likely GE members in the abundance diagram exhibits a wide range of vertical
angular momentum larger than what expected for the GE debris (i.e. \cite{koppelman2019}), in both prograde and retrograde
motion. This is likely due to contamination from other, smoother parts of the stellar halo. These properties are consistent
with what found for the SDSS DR7 calibration stars sample, and can be then considered as general.

\begin{figure*}[hbt!]
\centering
\vspace*{-0.4cm}
\includegraphics[width=160mm]{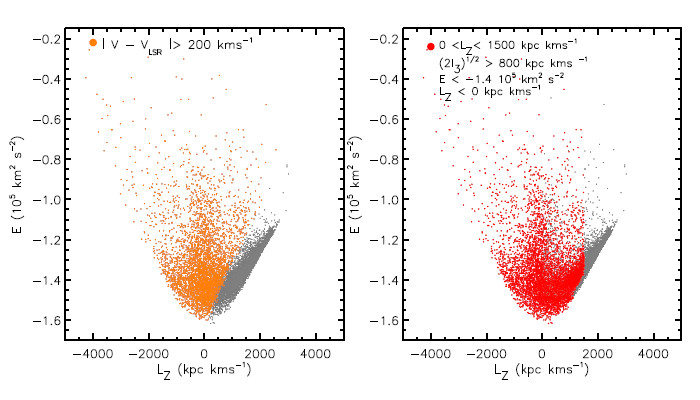}
\caption{
$E$ vs $L_z$ diagram for the SDSS DR7 calibration stars (grey dots). In the left panel, the orange dot symbols show the
distribution of stars selected by applying the high-velocity cut for the 3D velocities such that,
$|V_{\rm star} - V_{\rm LSR}| > 200$ km~s$^{-1}$. In the right panel, the red dot symbols represent the distribution of stars
selected by applying the new cuts in the phase-space defined by ($E$,$L_z$,$I_3$),
namely (1) $0 < L_z < 1500$ \kpckms, \I $> 800$ \kpckms \& $E < -1.5 \times 10^5$ km$^2$~s$^{-2}$, and (2) $L_z < 0$ \kpckms. We chose the more relaxed cut of \I $= 800$ \kpckms to take into account the overlap between stellar populations. 
}
\label{fig: halo_selection}
\end{figure*}

\section{Discussion}

\subsection{The stellar halo defined in the phase space}

Mapping our stellar halo while living inside it is not a straightforward task, because the information available for each
star is incomplete, even in the {\it Gaia} era. Indeed, the precise trigonometric parallaxes and proper motions are
available only in the confined local volume, and the line-of-sight velocities and metal abundances, derived from the
spectroscopic data (mostly provided by ground-based observations), are limited to the relatively bright stars in the
targeted regions of the sky. However, the advantageous position of the Sun in the Milky Way allows us to capture and
analyse most types of stellar orbits passing near the Sun. These orbits represent the basic ingredients of the stellar
halo \citep{may1986}.  Thus, unless the full-depth, full-sky astrometric and spectroscopic data of stars over the very
extended halo region are available, fundamental insights into this old and very important Galaxy component, can be
obtained primarily from the orbit-based selection in the phase space defined by the integrals of motion, $(E, L_z, I_3)$. 

Our analyses of the SDSS-SEGUE DR7 and APOGEE DR16 data sets suggest that halo stars are characterized by two distinctive
properties in the phase space: large values of the third integral of motion, $I_3$ [expressed in terms of \I],
such that \I $\gtrsim 1000$ \kpckms, and vertical angular momentum, $L_z \lesssim 1500$ \kpckms. Additionally,
as will be discussed later in Section 4.4.2, an important portion of halo stars populate the energy range below
$-1.5 \times 10^5$ km$^2$~s$^{-2}$. This energy threshold coincides (nearly) with the gravitational binding energy
at the position of the Sun, $\Phi(R=R_{\odot})$, as estimated from the observed escape velocity of nearby stars of
$V_{\rm esc} \sim 550$ km~s$^{-1}$ \citep[e.g.][]{sakamoto2003,koppelman2020},
so that $\Phi(R_{\odot}) = - V_{\rm esc}^2 / 2 \sim -1.5 \times 10^5$ km$^2$~s$^{-2}$.
These stars possess values of $I_3$ in the range of $500 <$ \I $< 1000$ \kpckms, which
corresponds to $\theta_{\rm orb} \sim$ 5-15~deg and, at the lower end of this interval of orbital angles (5~deg),
there exist contamination from the MWTD.

These general properties suggest that the selection of halo stars can be done by adopting
the fiducial limits of (1) 0 $ < L_z < 1500$ \kpckms \& \I $> 1000$ \kpckms \ ($\theta_{\rm orb} \gtrsim 15 - 20$~deg) \& $E < -1.5 \times 10^5$ km$^2$~s$^{-2}$, and (2) $L_z < $ 0 \kpckms. For retrograde halo stars ($L_z <$ 0 \kpckms), there is no need to apply a cut in \I. A more relaxed limit of \I $> 800$ \kpckms \ ($\theta_{\rm orb} \gtrsim 12$~deg), should be adopted to take into account the overlap between stellar populations and the uncertainties in the orbital parameters. In such a case, the selection will include some contaminants from the disk components, the MWTD in particular.

Note that $\theta_{\rm orb} \sim 15$~deg corresponds to $z \gtrsim 2.3$~kpc at the solar radius. It is also important to
remark that the suggested range of $L_z$ for halo stars ($L_z < 1500$ \kpckms), in the intermediate
metallicity interval of $-1.5<$ [Fe/H]$> -1$, contains also a fraction of MWTD stars.

To highlight the significance of these selection criteria for halo stars, we show, in Figure~\ref{fig: halo_selection},
the comparison between two $E$ vs. $L_z$ diagrams obtained by adopting the high-velocity cut
selection (left panel: $|V_{\rm star} - V_{\rm LSR}| > 200$ km~s$^{-1}$), and the new orbits-based selection criteria
in the ($E$, $L_z$, $I_3$) phase-space (right panel), for the case of the SDSS DR7 calibration stars.
The grey dots show the distribution for the entire sample, while the color-coded symbols represent stars selected by
employing the two methods. Examination of this figure reveals that the high-velocity cut selection
for halo stars misses a significant portion of the stellar halo represented by prograde stars with both, low and
intermediate energy, over the range of $0 \lesssim L_z \lesssim 1500$ \kpckms \ and
$E \lesssim -1.2 \times 10^5$ km$^2$~s$^{-2}$, whereas our selection criteria include these stars (right panel).
This clearly demonstrates the importance of an orbit-based selection, rather than a simple high-velocity cut, to avoid
a biased understanding of the Milky Way's stellar halo.

\subsection{Metallicity distribution and the halo duality}

Before {\it Gaia}, \citet{carollo2007,carollo2010} provided evidence of the Milky Way's halo complexity by demonstrating
that it comprises {\it at least} two components, namely, the inner and outer halo. The inner halo is characterized by a
mean Galactocentric rotational velocity $\sim$ 0, and a metallicity distribution function (MDF) with peak at
[Fe/H]$\sim -1.6$, while the outer halo exhibits highly retrograde motion and it is very metal poor, with an MDF having
peak at [Fe/H]$\sim -2.2$. The astrometric parameters used in the above analyses were not as accurate as those provided
by {\it Gaia} DR2, implying that it was possible to capture mainly the \enquote{coarse-grained} properties of halo stars. 

\begin{figure}[hbt!]
\includegraphics[width=80mm]{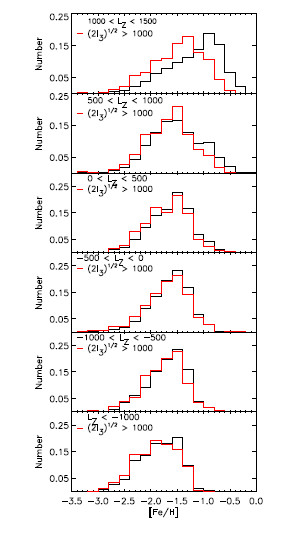}
\caption{
Metallicity distributions (MDFs) for the SDSS-SEGUE DR7 calibration stars in different ranges of $L_z$.
The black histograms represent the MDFs for the subsamples of stars selected in the various ranges of $L_z$,
starting from the top, $1000 < L_z = 1500$ \kpckms, to the bottom, $L_z < -1000$ \kpckms, while the
red histograms show the MDFs in the same ranges of $L_z$, but
reaching large distances from the Galactic plane during their
orbits, i.e., \I $> 1000$ \kpckms.
}
\label{fig: MDFs}
\end{figure}

To get further insights into this topic, we show, in Figure~\ref{fig: MDFs}, the MDFs
for the SDSS-SEGUE DR7 calibration stars determined for different values of $L_z$. The black histograms represent the MDFs for
the sub-samples of stars selected in various ranges of $L_z$, starting from $L_z$ = 1500 \kpckms, while the red histograms show
the MDFs in the same ranges of $L_z$, and reaching large distances from the Galactic plane during their orbits, i.e.
\I $> 1000$ \kpckms. We chose the more conservative cut in \I to avoid disk populations contamination (mostly MWTD).

In the range of angular momentum, $1000 < L_z < 1500$ \kpckms \ (top panel), the black histogram is dominated by the MWTD stellar
population, with metallicity peak at [Fe/H]$\sim -0.9$ ($L_z \sim$ 1200 \kpckms for the MWTD, \citealt{carollo2019}), and some thick
disk contamination. On the contrary, the red histogram shows lack of stars with disk-like kinematics, due to the selection over
larger values of \I, (\I $> 1000$ \kpckms), and exhibits a metallicity peak of [Fe/H]$\sim -1.3$. 
It is important to note that at larger \I, there exist a finite fraction of metal-rich halo stars with [Fe/H]$>-1$:
these stars are characterized by large $L_z$, $I_3$, and also $E$, and therefore high orbital eccentricities
\citep[see also,][]{bonaca2017, naidu2020}. Thus, this sub-sample (red color)  contains also a fraction of metal-rich stars with halo-like
kinematics, represented by the portion at the right-end of the MDF.

As $L_z$ decreases, in the range $500 < L_z < 1000$ \kpckms, the peak of the MDF represented by the black histogram is shifted
to a more metal-poor value of [Fe/H]$\sim -1.5$, while the fraction of MWTD stars decreases. 
Below $L_z = 500$ \kpckms the MDF shows a peak at [Fe/H] $\sim -1.5$, with a long tails toward
lower metallicities. At high \I, the MDF shows a dominant peak at [Fe/H] $\sim -$1.5, lack of stars with disk kinematic, and a large fraction of very metal-poor stars (second and third panel). In the ranges, $-500 < L_z < 0$ \kpckms, and $-1000 < L_z < -500$ \kpckms, the MDF is still dominated by stars with metallicity, [Fe/H] $\sim -$1.5 both, at low and high ranges of \I.

Note that, if we consider the mean rotational velocity and its dispersion determined for the inner halo component in \citet{carollo2010}, i.e. $V_{\phi} \sim$ 0 km~s$_{-1}$, and $\sigma_{V_{\phi}}$ = 90 km~s$^{-1}$, which correspond to $L_z \sim$ 0 \kpckms, and $\sigma_{L_z}$ = 800 \kpckms, then the range  $-1000 < L_z < 1000$ \kpckms, and high \I, samples very well the inner halo MDF (panels 2-4). 
Thus, the inner halo component includes the GE structure (dominant, around the peak of [Fe/H] $\sim -$1.5), low energy stars ($E \lesssim -1.5 \times 10^5$ km$^2$~s$^{-2}$), at [Fe/H] $< -$1, and metal-poor prograde stars, including the more metal rich, which are missed
by the cuts adopted in \citet{helmi2018}. 

When the stellar orbits become more retrograde, with $L_z < -1000$ \kpckms, the MDF progressively
includes more stars with [Fe/H]$< -2.0$. At high \I, the peak at [Fe/H] $\sim -$1.5 weakens, and shifts toward the very metal poor zone, at $\sim -$2. Such distribution is dominated by the outer halo population described in \citet{carollo2007,carollo2010}.\\

At what Galactic location the outer halo component starts to dominate?

\begin{figure*}[hbt!]
\centering
\vspace{-3.5cm}
\includegraphics[angle=180,width=170mm]{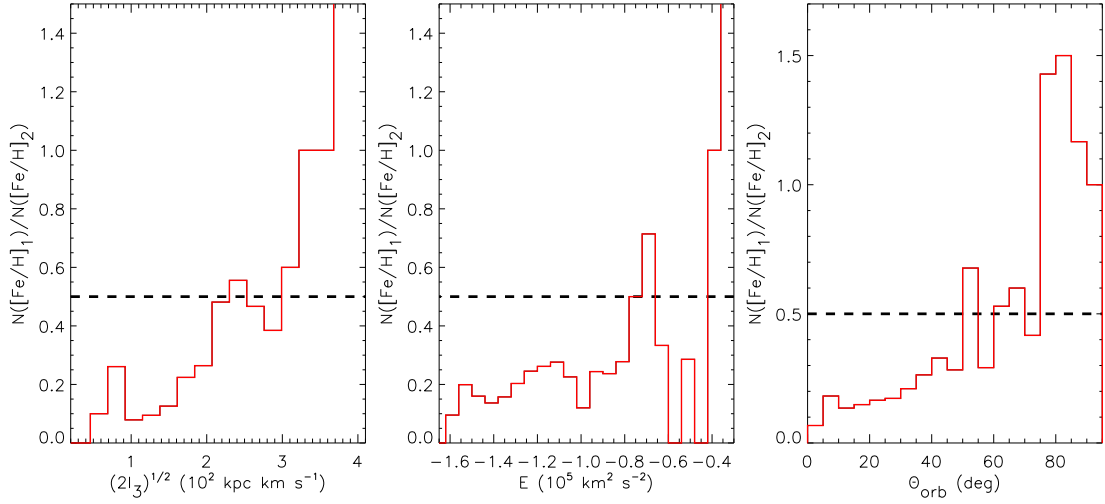}
\caption{
The ratio between the number of stars in the metallicity interval of [Fe/H]$<-2.2$, and that in $-2 <$[Fe/H]$< -1.4$,
$F = N({\rm [Fe/H]} < -2.2) / N(-2 <{\rm [Fe/H]}< -1.4)$, in each bin of
\I \ (left), $E$ (middle), and $\theta_{\rm orb}$ (right panel), for the SDSS DR7 calibration stars.
For the bins where $N(-2 <{\rm [Fe/H]}< -1.4)$ is zero ($F$ is infinity), occurring at some values of \I and $E$,
we set $F$ beyond the maximum of the adopted ordinate.
}
\label{fig: innner_outer_halos}
\end{figure*}

Figure~\ref{fig: innner_outer_halos} shows the ratio between the number of stars in the metallicity interval of [Fe/H]$< -2.2$, 
and that in $-2 <$[Fe/H]$< -1.4$, $F = N({\rm [Fe/H]} < -2.2) / N(-2 <{\rm [Fe/H]}< -1.4)$,
in each bin of \I \ (left), $E$ (middle), and $\theta_{\rm orb}$ (right),
for the SDSS-SEGUE DR7 calibration stars. The first range of metallicity represents
very metal-poor halo stars, where the outer halo is dominant with respect to the inner halo, which is well represented
in the second metallicity range ([Fe/H]$_{peak} \simeq -$1.6). Visual inspection of the left and right panels reveals that the relative fraction of
the very metal-poor stars increases discontinuously near \I $\simeq$ 2000 \kpckms \ and $\theta_{\rm orb} \simeq 50$ deg,
where $F$ = 0.5. This orbital angle corresponds to $z \simeq 10$~kpc at the solar radius. The ratio, $F$, becomes 1 or infinity (number of the stars in $-2 <$[Fe/H]$< -1.4$ = 0), beyond \I $\simeq 3000$ \kpckms, or
$\theta_{\rm orb} \simeq 80$~deg, which corresponds to $z \simeq 50$~kpc, at the solar radius. We remind here that $\theta_{orb}$ represents the maximum orbital angle from the Galactic plane.
The middle panel shows that outer halo stars possess very high energy, beyond $E \simeq -0.8 \times 10^5$ km$^2$~s$^{-2}$.


It is important to note that, while GE is still present in the metal-poor outer halo, it does not represent a significant
fraction of it (bottom panel in Figure~\ref{fig: MDFs}). Some of the retrograde sub-structures reported in recent literature, such as Sequoia \citep{myeong2019},
Arjuna and I'itoi \citep{naidu2020}, and the dynamically tagged groups (DTG) identified in \citep{yuan2020}, represent
\enquote{fine-grained} elements of the outer halo defined in \citet{carollo2007, carollo2010}. Thus, it is likely that the outer halo component is made of a superposition of several sub-structures, relics of past accretion events. This can be easily recognizable
in Figure 6, where the retrograde halo in the $E$ vs. $L_z$ diagram, resides in the area where several substructures have been identified, around $L_z \sim-$ 1000 \kpckms (see also \citealt{naidu2020}). \\ In addition, the SDSS-SEGUE calibration stars are selected to have the best atmospheric parameters and metallicity. This implies that
the MDFs of these stars represent the true distributions, in particular, at low metallicity, and for stars outside the disk
populations, and it is definitely not biased.

Also, in case of the inner halo stellar population,
the anisotropy parameter was found to be, $\beta$ = 0.6\footnote{$\beta$ is defined as $\beta = 1 -
(\sigma_t/\sigma_r)^{2}$, where $\sigma_r$ and $\sigma_t$ are
the radial and tangential velocity dispersions, respectively.} in \citet{carollo2010}, implying that the velocity ellipsoid is
radially elongated. This is in agreement with the radial anisotropy of the velocity ellipsoid determined for GE in
\citet{belokurov2018}, but this velocity anisotropy is now more significant, $\beta \sim 0.9$, after {\it Gaia}.  

We note that the asymmetry in the $E$ vs. $L_z$ distribution of halo stars, with the presence of more high-$E$ retrograde than prograde
stars, as seen in our and other local samples \citep{helmi2017,myeong2018a,myeong2018b,yuan2020}, is not caused by a selection effect that avoid contamination from stars with disk-like kinematics, mentioned by
\citet{naidu2020}. In our analysis, we have not adopted such an artificial selection method for halo stars and, as we have
also stated in Section 2: kinematically selected halo stars are generally biased in favor of high relative velocities with
respect to the Sun, giving asymmetry in the $E$ vs. $L_z$ distribution, but this is not the case in our work.

Also, as seen in Figure~\ref{fig: phase1}, largely retrograde stars with $L_z < -2000$ \kpckms \ and 
high $E$ of, say $E > -0.8 \times 10^5$ km$^2$~s$^{-2}$, possess values of the third integral of motion,
\I $\lesssim 1500$ \kpckms \ (or $\theta_{\rm orb} \lesssim 45$~deg). These stars may not be present in the H3 survey areas
of \citet{naidu2020}, where stars are selected at high Galactic latitudes $|b| > 40$~deg (see their Figure~1). This can explain
the lack of  asymmetry in the distribution of their sample. Moreover, while stars with such high negative $L_z$ possess large
negative $V_\phi$ in the solar neighborhood, they have small $|V_\phi|$ at large apocentric radii, due to angular-momentum
conservation, thus their rotational signals expressed in terms of $V_\phi$ is weaker in the outer-halo region (at large distance).
For instance, stars that possess $V_{\phi} \sim -200$ \kpckms at the location of the Sun will have
$V_{\phi} \sim -20$ \kpckms, at $R = 100$ kpc.

\begin{figure*}[hbt!]
\vspace{-6cm}
\centering
\includegraphics[angle=90, width=170mm]{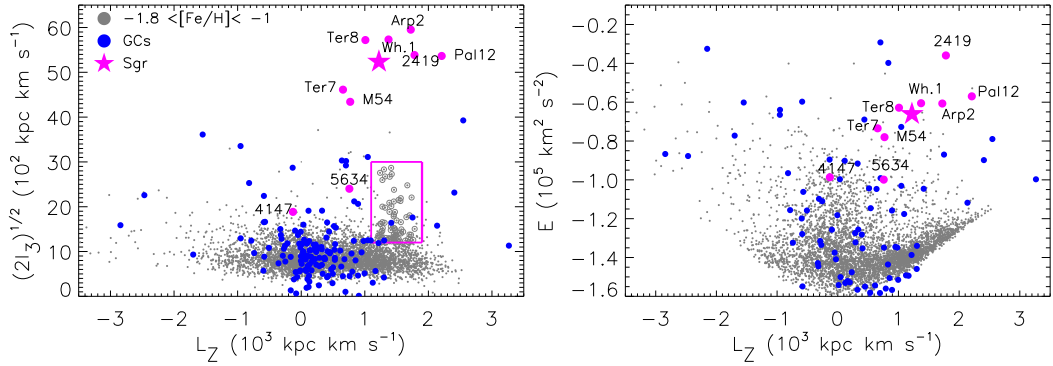}
\caption{Sagittarius dwarf galaxy (pink star) and the Milky Way globular clusters (pink circles with names labelled, and blue
circles), together with the SDSS DR7 calibration stars in $-1.8<$[Fe/H]$<-1$ (grey dots), in the \I \ vs. $L_z$ (left panel),
and $E$ vs. $L_z$ diagrams (right panel).
In the left panel, stars in the region where the 'trail' feature is identified are shown with filled circles inside
a pink box.
}
\label{fig: Sgr}
\end{figure*}

\subsection{On the relation with the Sagittarius dwarf galaxy, its stream, and the Milky Way's globular clusters}

As clearly shown in Figure \ref{fig: phase1} (left panels), the $L_z$ distribution of
halo stars in the metal-poor ranges of [Fe/H]$<-1.4$ is sharply bounded at $L_z \simeq 1500$ \kpckms. 
This appears to be the case for both SDSS DR7 and DR16, and
 there exist no selection effects to cause this discontinuous distribution of stars
(See Figure~\ref{fig: phase_selection}).
Therefore, the truncation of $L_z$ at this value can be considered a universal property for candidate halo stars in the
local volume. Also, as mentioned in Section 3.1.2., in the SDSS DR7, where the halo system is better represented than in
the APOGEE DR16, 
there exist an elongated structure in the \I \ vs. $L_z$ diagram for the metallicity ranges of $-2.2 <$[Fe/H]$< -1.4$,
namely from $(L_z, (2I_3)^{1/2}) \simeq (1500-1800, 400)$ \kpckms to $(1000, 3500)$ \kpckms.
We note that the truncation value of the vertical angular momentum below [Fe/H] $< -$1.4, $L_z \simeq 1500$ \kpckms,
is just adjacent to the lower-end of this elongated structure located at $L_z \simeq 1500-1800$ \kpckms, thus, these two
features may be related to each other.

To get more insights into this issue, we explore a possible connection between the sharp cut in $L_z$ of halo stars with the Sagittarius dwarf spheroidal galaxy (Sgr
dSph), its associated stream (Sgr stream), and the Milky Way's globular clusters (GCs), by inspecting their phase-space distribution.

This is motivated by the fact that the Sgr stream contains metal-poor stars over
the range of $-1.9 \lesssim$[Fe/H]$\lesssim -1$, with mildly peak at around [Fe/H]$= -1.4$ \citep[e.g.,][]{deboer2015,
hayes2020}. Such metallicity interval overlaps with the range where the elongated features is spotted in the
left panels of Figure~\ref{fig: phase1} (\I \ vs. $L_z$ diagram). 


For the Sgr dSph, we adopt the recent compilation reported in \citet{pawlowski2020} (see their Table 1) based on the works
by \citet{ibata1997}, \citet{dinescu2005} and the \citet{gaia2018}. For the Milky Way's GCs, we adopt the recent data collection 
described in \citet{baumgardt2019}, where {\it Gaia} DR2 proper motions are included.
Figure~\ref{fig: Sgr} shows the \I \ vs. $L_z$ and $E$ vs. $L_z$ diagrams for the Sgr dSph and the Milky Way's GCs superimposed 
with the distribution
of the SDSS-SEGUE DR7 calibration stars (grey dots) in the range of metallicity, $-1.8 <$[Fe/H]$< -1$, where the elongated 
feature is more evident.  

Examination of Figure~\ref{fig: Sgr} reveals that the GCs M~54, Ter~7, Ter~8, and Arp~2 are very close to Sgr dSph in the phase
space $(E,L_z,I_3)$, indicating that they are associated to the dwarf galaxy. This is in agreement with previous results
reported in \citep[e.g.,][]{law2010, bellazzini2020}, where the association was based on the GCs position and velocity.
Figure~\ref{fig: Sgr} also shows that Whiting~1 and Pal~12, are also close to the location of the Sgr dSph
in the $(E,L_z,I_3)$ phase space. These GCs are actually found in the Sgr trailing arm, and therefore, they reflect the
orbital motion of the Sgr dSph \citep{bellazzini2020}.

NGC~5634, NGC~4147 and NGC~2419, which may be associated with more ancient wraps of the stream \citep{bellazzini2020},
have very different $I_3$ values (except NGC~2419), with respect to the Sgr dSph, perhaps caused by the large deviation of the
stream from the galaxy's orbit. We also notice that the positions of the Sgr dSph and its associated GCs in the $E$ vs. $L_z$
diagram are in good agreement with the Sgr-associated substructures of field stars found in \citet{naidu2020}. 

It is important to remark that the $L_z$ value of the Sgr dSph ($\simeq 1200$ \kpckms) is close to the sharp bound found
in the \I \ vs. $L_z$ distribution for [Fe/H]$\lesssim -1.4$, at $L_z \sim 1400-1500$ \kpckms, and the mentioned elongated
structure of stars in the \I \ vs. $L_z$ diagram is somehow oriented from
this truncation region. Moreover, the Sgr dSph and some GCs are distributed along this structure. 
Thus, the sharp $L_z$ boundary, and the elongated feature in the ($L_z$, \I) diagram of local halo stars, 
might be driven by a perturbation from the Sgr dSph in some collective manner. A similar dynamical effect from
the Large Magellanic Cloud (LMC) was recently examined by \citet{garavito2019}, and \citet{cunningham2020}, motivated by a recent
determination of the LMC mass, of the order of $10^{11} M_{\odot}$  \citep[e.g.,][]{erkal2019}.
This massive LMC can produce a dynamical effect on both dark-matter and stellar halos of the Milky Way.
In such a scenario, the infalling LMC to the Milky Way generates a pronounced wake to the distributions of both dark-matter
particles and halo stars. The wake is decomposed in transient and collective responses, where the former (latter) effect
is local (global), and provides distinct kinematic patterns in both dark-matter and stellar halos.
Similarly to the LMC wake-effect, we suggest that an equivalent but weaker wake is induced by the Sgr dSph in the local
halo stars, although further theoretical investigation is needed to confirm this hypothesis.

The Sgr dSph is also thought to affect the structure and dynamics of the Milky Way disk \citep[e.g.,][]{purcell2011,
laporte2018}. In particular, the galaxy can produce a wobbly structure in the outer parts of the disk \citep{laporte2018}
as found in recent works \citep{li2017,bergemann2018}. 
Thus, the Sgr dSph may play an important role in forging the dynamical structures in the Milky Way, and in its star formation history \citep{ruizlara2020}.

It is also interesting to note in Figure~\ref{fig: Sgr} that some GCs occupy the phase-space
region of the GE debris stars, around $L_z = 0$, in both the \I \ vs. $L_z$ and $E$ vs. $L_z$ diagrams. 
This suggest that these GCs, and field halo stars, in this phase-space region, originate from the same progenitor galaxy
that formed the GE structure. Also, a fraction of field halo stars in this phase-space region may originate from
disrupted GCs \citep[See, e.g.,][]{massari2019, kruijssen2019, myeong2019, kruijssen2020, forbes2020}.

\subsection{What is the in situ stellar halo?}

\subsubsection{Formation process of dark-matter and stellar halos}

The definition of the so-called in-situ stellar halo, compared to the ex-situ stellar halo, and other halo populations, is 
somewhat different in various literatures, in particular, among observational studies
\citep[e.g.,][]{sheffield2012,hayes2018,haywood2018,dimatteo2019,gallart2019,conroy2019,montalban2020,belokurov2020,naidu2020}.
This is due to the numerous sources of halo stars generated during the complex galaxy formation process, and the
difficulty in identifying such stars in observational studies. Theoretical predictions suggest that in-situ stars formed
inside a parent halo within the virial radius \citep{zolotov2010,tissera2012,tissera2013}.

While dark halos are entirely formed through hierarchical assembly and merging process, stellar components
in dark halos originate from multiple sources, in particular, from star formation within cooled gas originally present in
parent halos, in gas stripped from merging satellites and subsequently cooled, 
or stars are supplied by merging/accreting halos, that disappear after tidal interaction, or survive as luminous satellites,
\citep[e.g.,][]{bekki2001,bullock2005,zolotov2009,purcell2010,font2011,mccarthy2012,tissera2013,cooper2015,rodriguez2016,deason2016,dsouza2018a,monachesi2019,fattahi2020}.
The question is whether we can really distinguish stars formed inside parent halos from those supplied
from outside, when using observational data alone.

A brief review of the halos formation process in the expanding Universe can provide some insights into this issue.

Formation of dark halos in the expanding Universe is totally hierarchical \citep{white1978}:
in the standard structure formation scenario
based on a $\Lambda$-dominated cold dark matter (CDM) models, small, dense dark halos collapsed first while the Universe was
dense, and they merged together leading to larger, more massive halos. Subsequent collapsed dark halos are generally less
dense as they formed in a less dense Universe. The merging process between the early high density small halos, and the less
dense large halos, results in a more massive halo, where its central compact part is dominated by
the denser halo progenitor, while its outer and lower-density section, is made of the larger halo progenitor debris.
In this hierarchical process, the {\it main progenitor halo} can be defined as the most massive parent halo, when following
backward the branching of a merger tree for a current host halo \citep[e.g.,][]{mo2010}.

The formation and evolution of stellar halos differ from those of dark halos. Indeed, stars are formed
from cooled interstellar gas, supplied after radiative cooling of virialized hot gas, mostly at the bottom of each dark halo  
\citep{rees1977,fall1980}, or from already cold gas directly flowing from intergalactic space \citep{dekel2009}.
Gas in each halo is also supplied by merging/accreting satellites carrying cold gas cores, and hot halo gas
\citep[e.g.,][]{kauffmann1993}. 
Each dark halo grows through merging/accretion with other dark halos, and stellar populations within the product of this
merging, are then composed of pre-existing stars formed inside the main progenitor halo, and merged/accreteted stars
in other halos. The former stellar population formed inside a main parent halo can be regarded as an in situ stellar halo,
which consists of stars forged from cooled gas inside it, and those dynamically heated from an already formed stellar
disk in the merging process of subhalos/satellites
\citep[e.g.,][]{zolotov2009,purcell2010,font2011,mccarthy2012,tissera2013,cooper2015,rodriguez2016,monachesi2019}.
Detailed simulation studies for the formation of stellar halos suggest that the fraction of in-situ relative to
ex-situ stellar halos increases with the increasing total mass of a main progenitor halo \citep{rodriguez2016,monachesi2019}.

\begin{figure*}[hbt!]
\centering
\includegraphics[width=120mm]{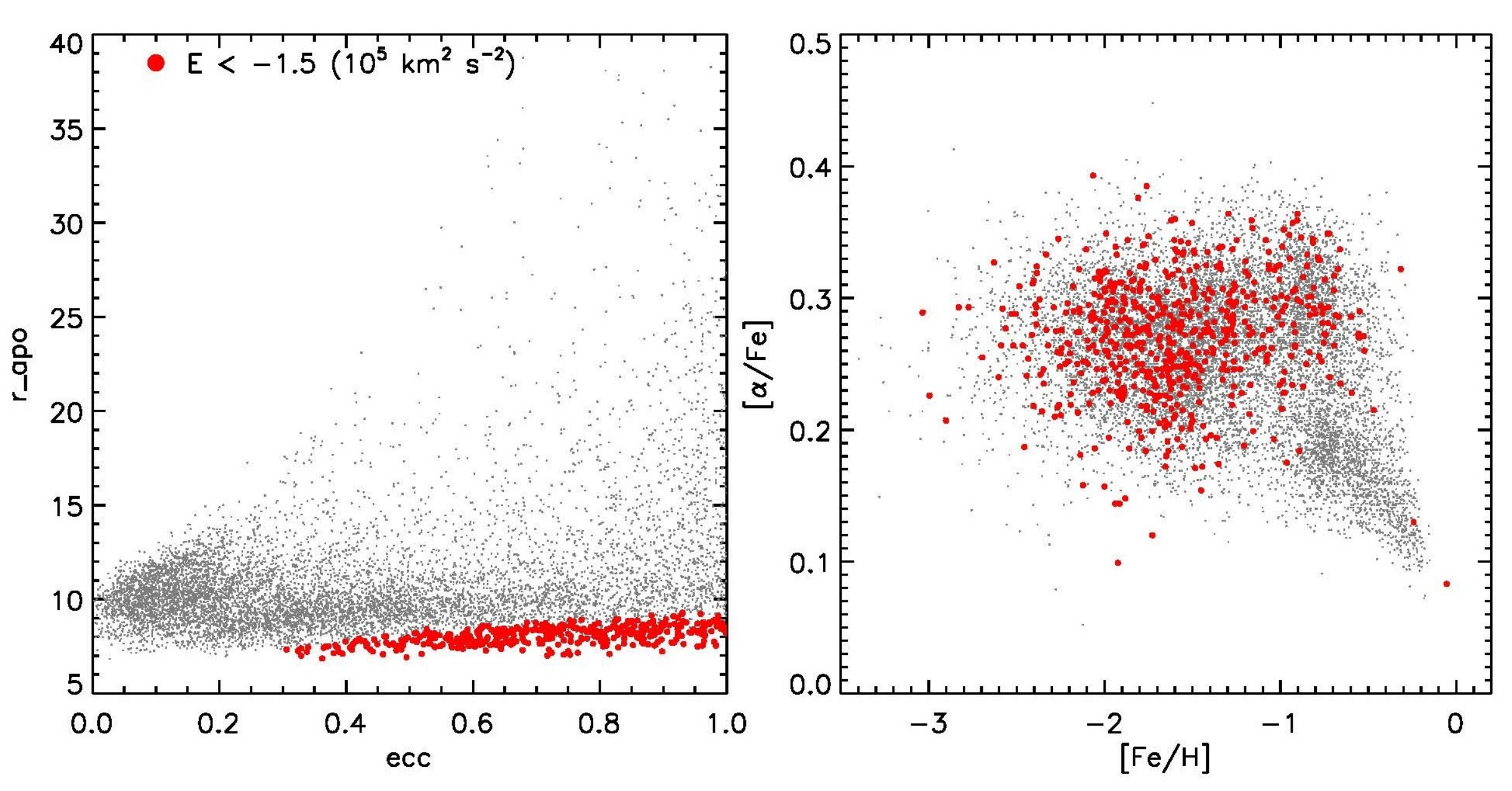}
\caption{The properties of the lowest binding energy stars ($E < -1.5 \times 10^5$ km$^2$~s$^{-2}$; red circles)
for the SDSS-SEGUE calibrations stars, selected as candidate in situ halo stars,
in the apocentric radii vs. orbital eccentricities (left panel) and the [$\alpha$/Fe] vs. [Fe/H] diagrams (right panel).
Grey dots show the entire sample.
Notably, these stars are orbiting inside the solar circle, $r < R_{\odot}$, having a vast range of orbital eccentricities
of $e > 0.3$. The right panel shows that these lowest $E$ stars are widely distributed in the [$\alpha$/Fe] vs. [Fe/H] diagram
below [Fe/H]$= -1$, and also, in the region of $-1 \lesssim$[Fe/H]$\lesssim -0.8$, and [$\alpha$/Fe]$> 0.25$.
}
\label{fig: lowest_E}
\end{figure*}

\subsubsection{Lowest binding energy stars as a part of the in situ stellar halo}

How do we select stars belonging to the in-situ stellar halo, from observational data in the local volume?
This has been a central theme in recent related works
\citep[e.g.,][]{hayes2018,haywood2018,dimatteo2019,gallart2019,conroy2019,montalban2020,belokurov2020,naidu2020}.

A finite fraction of stars, formed from cooled gas in the bottom of a main progenitor halo, or from
merged/accreted cold gas supplied by other halos, are most tightly bound to the gravitational
potential of that progenitor halo after dissipative cooling. In comparison, other in-situ stars, which are product of
a pre-existing dynamically heated disk by merging events, and ex-situ stars supplied by merging/accretion events
from outside, are less bound, because of their larger binding energies.
Then, candidates in-situ halo stars can be defined as those being most tightly bound to the Milky Way gravitational
potential, or, in other words, those having the lowest binding energy, $E$.
As mentioned above, another possible source of in-situ halo stars, are those kicked out from a pre-exiting,
metal-rich high-$\alpha$ disk populations, driven by merging events with satellites or subhalos and are expected to
have higher $E$ after dynamical heating \citep[e.g.,][]{bonaca2017,belokurov2020}. 

Motivated by these scenarios, we consider stars with
$E < -1.5 \times 10^5$ km$^2$~s$^{-2}$ as candidate in-situ halo stars. This choice can be understood by examining the properties of halo stars in the DR7 sample.
The bottom-middle panel of Figure~\ref{fig: phase2_sub} and \ref{fig: DR16_phase2_sub} shows that the extension of candidate GE debris stars, color-coded with dark blue, have energy above $-1.5 \times 10^5$ km$^2$~s$^{-2}$ at $L_z \sim 0$. Thus, stars selected below this energy value should be less contaminated by those associated with the GE structure.
This is in agreement with the results of \citet{naidu2020}, where the GE debris stars, selected in terms of orbital
eccentricities ($e >$ 0.7), occupy only a small fraction at energies, $E < -1.5 \times 10^5$ km$^2$~s$^{-2}$,
in the $E$ vs. $L_z$ diagram (see their Figure 11). A natural explanation for this effect reside in the formation mechanism of the GE debris, which was likely formed by the merging of a dwarf galaxy with associated dynamical heating of a pre-existing stellar disk. It is expected that these stars possess high binding energy after the merging event.

As discussed in Section 3.1.1, stars with the lowest $E$ range are distributed around $L_z = 0$, and
populate the third integral of motion in the range of $400 \lesssim$ \I $\lesssim 1200$ \kpckms. Stars with the lowest $E$
have \I $\simeq 700$ \kpckms (see right panels in Figure~\ref{fig: phase1}). These value of \I correspond to orbital angles
in the range of $5^\circ \lesssim \theta_{\rm orb} \lesssim 20^\circ-30^\circ$, and $\theta_{\rm orb} \simeq 10^\circ$
for the lowest $E$. When expressed in terms of the maximum distance from the Galactic plane, $z_{\rm max}$,
these stars have $z_{\rm max} < 5$~kpc.

Figure ~\ref{fig: lowest_E} shows the apocentric radii, $r_{\rm apo}$, as a function of the orbital eccentricity, for the DR7 calibration stars sample (grey dots). The red dots represent stars with $E < -1.5 \times 10^5$ km$^2$~s$^{-2}$, and their $r_{\rm apo}$ are near but below the solar position,
$R_{\odot}$, and their orbital eccentricities are distributed over the range, $e>0.3$. A fraction of these stars may be
contaminated with GE debris, at $e>0.7$, although such high $e$ stars can actually be in-situ halo stars in the inner region of the Milky Way. The right panel of Figure~\ref{fig: lowest_E} shows the [$\alpha$/Fe] vs. [Fe/H] diagram for the DR7 calibration stars (grey), and the candidate in-situ stars (red). The majority of the low-energy stars occupy the region at [Fe/H]$< -1$, but some of them are also in the area dominated by the MWTD, with small contamination from the thick disk ([Fe/H]$\lesssim -0.8$ and [$\alpha$/Fe]$> 0.25$).
As can be seen in the top-middle and top-right panels of Figure~\ref{fig: phase2_sub}, disk-like stars are located just around the adopted threshold of $E = -1.5 \times 10^5$ km$^2$~s$^{-2}$, in the $E$ vs. $L_z$ diagram, therefore, the contamination from these stars can be removed by adopting a slightly smaller value of $E$.
Thus, candidate in-situ halo stars can be selected by adopting $E < -1.5 \times 10^5$ km$^2$~s$^{-2}$, and [Fe/H]$< -1$.

We argue that it is 
unlikely that these stars originate from the dynamical heating of a pre-existing high-[$\alpha$/Fe] stellar disk driven by
merging of subhalos or satellites at early epoch, nor from the debris of a merging event of a dwarf galaxy
giving rise to the GE structure.
Instead, these low-$E$ stellar populations may have been formed 
from early dissipation processes supplied by
cold-gas stream \citep{dekel2009}, where efficient star formation and chemical evolution are at work
in the bottom region of a host dark halo \citep[e.g.,][]{brook2020}.

\subsubsection{Other possible candidates for the in-situ stellar halo}

In-situ stars that are formed from infalling cooled gas within a dark halo are subject to chemical evolution,
accompanying not only gaseous infall but also outflow from the system, and the interplay between these hydrodynamical
processes yields a specific metallicity distribution of stars, where a finite fraction is metal-poor,
say [Fe/H]$<-2$ \citep[e.g.,][]{chiappini1997,chiappini2001,haywood2013,toyouchi2018}.
In particular, early infalling, pristine gas would have been settled into an equatorial plane of
a progenitor dark halo, in the presence of an initial angular momentum \citep[e.g.,][]{katz1991}.
From this gas, a fractions of very metal-poor stars, with low $L_{\perp}$,
and relatively high $L_z$ may have been formed and left there.

Do such metal-poor stars exist in the Milky Way?

In the sample of \citet{chiba2000}, which was assembled from several
stellar sources in combination with the {\it Hipparcos} satellite data, and ground-based observations \citep{beers2000}, a finite
fraction of metal-poor halo stars possess low $I_3$ (low $L_{\perp}$), low $J_\nu$, and high $L_z$, as shown in their
Figure~16. Note that these stars are located at relatively low Galactic latitudes (see Figure~1 of \citet{beers2000}), and therefore, stars with the above orbits are indeed present in a local sample. Also, in a recent work based on data from the Pristine Survey \citep[]{starkenburg2017}, \citet{sestito2020} report the discovery of several metal-poor stars, with metallicity, 
[Fe/H]$=-2.5$, low orbital eccentricities, and low vertical actions, $J_z$. These stars show both prograde and
retrograde rotations, where the former appears to occupy a larger fraction. The idea is that such stars formed,
possibly, in the early stages of the Milky Way assembly process, and may still be in the Galactic plane.

Some, or many of these stars, may acquire vertical actions through early dynamical processes, and thus
they exhibit a finite $I_3$ integral of motion, leaving a fan shape in the $E$ vs. \I diagram, as seen in the
metal-poor intervals of Figure~\ref{fig: phase1}.  Although some of these metal-poor stars have been identified
in recent surveys, their global properties are still not well understood.

In connection to this issue, it is interesting to remark that recent high-redshift surveys of star-forming galaxies,
based on carbon monoxide (CO) or H$\alpha$ spectroscopy, have identified rotating gaseous disks at redshifts
$z \sim 2$ \citep{forster2009,price2016,genzel2017}. Furthermore, in most recent studies, based on the Atacama Large
Millimetre/submillimetre Array (ALMA), galaxies selected through their [C~II] emission spectroscopy of H~I absorption,
show that such gaseous disks on a galaxy scale are already present at redshifts of 4 to 5, or 12 Gyrs ago
\citep{neelman2019,neeleman2020}. The existence of such disks at high redshift, suggests that the accretion of cold gas
onto a dark halo is at work \citep{dekel2009}, and explains the very early formation of rapidly rotating gaseous disks
at $z \sim 5$. The rotational properties of these disks are characterized by a flat rotation curve, thus suggesting
the presence of dark halos behind them. The existence of rotating gaseous disks naturally implies star formation activity,
leading to the formation of metal-poor stars with disk-like kinematics, like those found in the Pristine Survey
\citep{starkenburg2017,sestito2020}.

Note also that the rotational direction of such early gaseous disks, formed from cold accretion flow, is not necessarily
the same as that of the currently observed thin/thick disks at $z=0$. This implies that ancient metal-poor halo stars
formed in this way can be rotating in either way (prograde or retrograde) compared to the current thin/thick disks,
as observed in \citet{sestito2020}. These stars may occupy regions of the phase-space located along a parabola
in the $E$ vs. $L_z$ diagram (see Figure~\ref{fig: phase1}).
We suggest that many of the in-situ halo stars originated from the above process can be present in the local volume, however such stars are likely outside the survey volume of SDSS. 


\section{Implications for the formation of the stellar halo}

\subsection{A brief overview prior and subsequent to {\it Gaia}}


Since \citet{els1962} first presented their view of Galaxy formation in terms of a monolithic, free-fall collapse
scenario, several pieces of evidence, obtained from subsequent observational and theoretical studies, have suggested that
both processes, the dissipative cooling of gas and the dissipation-less merging of small stellar systems, are at work in the formation of
the Milky Way's stellar halo \citep[e.g.,][]{bekki2001}.
Recent compelling evidence of such \enquote{mixed} formation scenario was shown by \citet{carollo2007,carollo2010}, based on
their finding of a multiple-halo structure, namely, the inner- and outer-halo.
The inner halo, which corresponds to the flattened part of the halo
originally found by \citet{sommer1990}, was suggested to form through the dissipative merging of massive clumps, leading to
stars with high eccentric orbits. Part of this halo may have been formed in-situ from the rapid collapse of
infalling gas. On the contrary, the distinctive proprieties of the outer halo stellar population, in particular, its
retrograde-rotation and very metal-poor MDF, implied a different origin, possibly through a dissipational caothic merging of low-mass subsystems.
Later numerical simulations of hierarchical galaxy formation were able to reproduce these dual-halo systems, which exhibit
many of the observed properties of the Milky Way, as well as M31
\citep{zolotov2009,font2011,mccarthy2012,tissera2012,tissera2013,tissera2014}.

The halo age-structure described in \citet{santucci2015}, \citet{carollo2016}, and \citet{whitten2019}, have also added
significant evidence for the inside-out assembly of the halo, with the contribution of the latest merger events to the outskirts.
Interestingly, high-resolution numerical simulations in $\Lambda$CDM are able to reproduce the observed halo age gradient slopes by 
considering mainly the contribution of the accreted stars, with a minimal input of in-situ stars \citep{carollo2018}.

After {\it Gaia}, this view of hierarchical clustering scenario for the stellar halo formation has been much strengthened and
detailed, owing to the precision of astrometric data and, more importantly, the associated dramatic increase of 
stellar samples, of the order of a billion. Then, the radially anisotropic velocity field of halo stars in the intermediate
metallicity range of $-1.6<$[Fe/H]$<-1$, which corresponds to high-$e$ stars first interpreted as evidence of
free-fall Galaxy collapse by Eggen et al., and also a significant portion of the inner halo described in
Carollo et al., is actually found to be due to a coherent stellar population
in the $E$ vs. $L_z$ diagram \citep[Gaia-Enceladus:][]{helmi2018} as well as an elongated feature in the velocity space
\citep[Sausage:][]{belokurov2018}, quantified as $\beta \sim 0.9$. This structure can be a stellar remnant of an infalling,
massive dwarf galaxy  \citep{helmi2018,koppelman2018,deason2018,belokurov2018,haywood2018,myeong2018c}, probably driven by the
decay of its orbit through dynamical friction, i.e., loss of its orbital kinetic energy through long-range gravitational
interaction with surrounding dark-matter particles.

\subsection{Implications from this work}

The analysis of a large sample of stars in the phase-space defined by ($E$,$L_z$,$I_3$), and over the metallicity ranges
that include the GE structure (Figure~\ref{fig: histo_phase}), suggests that metal-poor halo stars show a nearly exponentially
decreasing distribution with increasing $E$ and $I_3$, at $E > -1.4 \times 10^5$ km$^2$~s$^{-2}$, and \I $> 800$ \kpckms.
This exponential dependence of a distribution function with $E$ is achieved in lowered isothermal models, i.e.,
a family of King models for stellar systems in dynamic equilibrium. There exist also a family of anisotropic spherical models
with an exponential dependence on $L$, which corresponds to $I_3$ in the current non-spherical case \citep{binney2008}.
As one of such anisotropic models, we have chosen here the Michie-Bodenheimer type distribution shown in
Equation~(\ref{eq: michie}) \citep[e.g.,][]{sommer1987} to reproduce the global distribution functions presented in
Figure~\ref{fig: histo_phase}, but any other similar models can be applied as well.

The exponentially decreasing trend with increasing $E$, is found over the metallicity ranges of
$-1.8<$ [Fe/H] $<-1.4$, and $-2.2<$ [Fe/H] $<-1.8$, which include not only the GE debris, but also metal-poor prograde stars,
missed by the high velocity cut selection. Such a trend suggests that this part of the halo is on the way, but perhaps close,
to the dynamically relaxed state like the family of King models, although there remain small-scale, yet unrelaxed substructures,
that make local deviations from a general exponential distribution. It is also evident that many of the stars with lower binding
energies of $E < -1.5 \times 10^5$ km$^2$~s$^{-2}$, which are candidate in-situ stars, have their apocentric radii below
the solar radius. This implies that such stars are largely underrepresented in local samples like the one adopted in this paper, as first suggested by
\citet{sommer1990} (See their Figure 5), and later by \citet{chiba2000} (their Figure 12). The lack of low-energy stars explains the drop of
the distribution functions below $E = -1.5 \times 10^5$ km$^2$~s$^{-2}$ in Figure~\ref{fig: histo_phase}. Future surveys extended 
over the inner regions of the Milky Way inside the solar radius, including SDSS-V \citep{kollmeier2017},
are expected to fill out this missing part of the distribution function at the lowest $E$ side.

Our analysis also suggests that the $I_3$ distribution of the GE structure is less extended than that of more metal-poor stars
,or equivalently, the GE structure exhibits lower maximum orbital angles, $\theta_{\rm orb}$, (Figure~\ref{fig: histo_phase} and \ref{fig: innner_outer_halos}). This may indicate that a progenitor dwarf galaxy is falling
into the main parent halo of the Milky Way at a finite angle from the equatorial plane, and the stellar debris are spread out in the
course of the decaying orbital angles.

The very metal-poor stellar halo, [Fe/H]$<-2.2$, consists of stars with both prograde and retrograde rotation.
In this range of metallicity, the $L_z$ distribution is asymmetric, with a peak at $L_z \simeq 0$, and an extension towards retrograde motions, $L_z < 0$
(See Figure~\ref{fig: histo_phase}); the preferential high retrograde motion for very metal-poor stars was identified
by \citet{carollo2007,carollo2010} before Gaia and later by \citet{helmi2017} and \citet{myeong2018d} with Gaia.
It is striking that these very metal-poor stars exhibit a more extended distribution toward larger \I \ or $\theta_{\rm orb}$,
with respect to the more meta-rich halo stars ($-$2 $<$ [Fe/H] $< -$1.4), which include the GE structure (Figure~\ref{fig: innner_outer_halos}). This property reveals the existence of two halo layers: a flattened inner halo (dominated by the GE structure), and a more extended outer halo \citep{sommer1990,chiba2000,carollo2007, carollo2010,belokurov2018}. 
The fact that the very metal poor outer halo possesses large \I values, and exhibit a {\it random spin} (prograde and
retrograde stars; left-bottom panel of Figure~\ref{fig: phase1}), suggests that it was likely originated from a chaotic,
dissipationless, and random merging, and accretion, of less massive dwarf satellite galaxies. Candidate surviving of such
low-mass systems are the ultra-faint dwarfs (UFDs), as envisaged in \citet{carollo2007}.

Thus, the outer layer of the halo is accretion dominated. Indeed, many of such low-mass satellites have probably fallen into
the Milky Way during the very early stage of Galaxy formation, as suggested by their long-term orbital evolution using {\it Gaia DR2}'s proper motions
\citep{miyoshi2020}. Some of the low-mass satellite galaxies have high $E$ and large negative (retrograde) $L_z$
\citep{myeong2018d}, and exhibit lower [$\alpha$/Fe] ratios than the GE structure \citep{matsuno2019}. This implies that
they have a longer star formation timescale and thus contain younger stars than those comprising the more flattened part of
the halo, dominated by the GE structure. Thus, the outer halo may comprise different layers as well: one made of accreted
very old and very metal poor low-mass galaxies (stars are in a more relaxed state, smooth outer halo), and one made of recent
accretion events of dwarf satellite galaxies, that are not relaxed yet.
It is interesting to note that the latter layer of the outer halo is in agreement with the reported spatial dependence
of the ages of globular clusters, where those in the outer parts of the halo, say at $r > 20$~kpc, are systematically younger
than in the inner parts \citep{zinn1985,mackey2004}. The existence of distinct layers in the outer halo is also confirmed by the age-structure of the Milky Way's stellar halo described in \citet{santucci2015} and \citet{carollo2016}. In the later analysis it was inferred a negative age gradient of $-$25 Myr kpc$^{-1}$ (in the radial distance direction), and it was shown, in addition to the existence of a very old central region up to 10-15 kpc, the presence of several younger structures in the outskirts of the halo.   

\section{Summary of the main results and prospects}

In this paper we explored the general properties of the Milky Way's halo system inferred from the analysis of a large
sample of stars in the phase-space defined by the three integrals of motion, ($E$, $L_z$, $I_3$), in combination with the
chemical abundances. $I_3$ is the so-called third integral of motion, which is analytically defined in a St\"akel-form
gravitational potential. We have especially focused on the \enquote{coarse-grained} phase-space distribution of halo stars, namely,
their averaged properties over the phase space being close to a dynamically steady state, rather than the \enquote{fine-grained} phase-space distribution, which includes not-yet relaxed small-scale substructures. The main results can be summarized as follows:

\begin{itemize}
\item We propose to select halo stars by using an orbit-based selection method, which does not introduce selection bias
associated with high velocity cuts. It is found that halo stars, which are distinguished from disk stars having
low orbital angles from the Galactic plane, can be selected
over the following two domains in the ($E$, $L_z$, $I_3$) phase-space:

\begin{eqnarray}
&{\rm (1)}& \ 0 < L_z < 1500 \ {\rm kpc}~{\rm km}~{\rm s}^{-1} \nonumber \\
            &  & \& \ (2I_3)^{1/2} > 1000 \ {\rm kpc}~{\rm km}~{\rm s}^{-1} \nonumber \\
            &  & \& \ E < -1.5 \times 10^5 \ {\rm km}^2~{\rm s}^{-2}\\
&{\rm (2)}& \ L_z < 0 \ {\rm kpc}~{\rm km}~{\rm s}^{-1} .
\label{eq: for_halo}
\end{eqnarray}

The cut, $(2I_3)^{1/2} > 1000$ \kpckms \ in (1) corresponds to an orbital angle of $\theta_{\rm orb} = 15 - 20$~deg.
A more relaxed limit of \I $> 800$ \kpckms \ ($\theta_{\rm orb} \gtrsim 12-15$~deg), should be adopted to take into account the overlap between stellar populations and the uncertainties in the orbital parameters. In such a case, the selection will include some contaminants from the disk components. The value of $E = -1.5 \times 10^5$ km$^2$~s$^{-2}$ in (1)
nearly coincides with the gravitational binding energy at the position of the Sun.

\item Stars with the lowest binding energy have a non-zero third integral of motion, expressed as \I $\simeq$ 700 \kpckms,
and $\theta_{\rm orb} \simeq 10$~deg. These stars have $L_z = 0$.

\item Stars with a large vertical angular momentum $L_z \simeq 2000$ \kpckms, which are mainly thin disk stars, possess
low $I_3$, expressed as \I $\lesssim$ 400 \kpckms, while stars with $L_z \le 2000$ \kpckms \ have always \I $>$ 400 \kpckms.
There are no stars having both $L_z \simeq 0$ and \I $= 0$, and this is not due to the selection effect of the current
local sample.

\item The metal-weak thick stars possess \I values larger than the thick disk, which implies larger orbital angles.
The average orbital angles result, $<\theta_{orb}> \sim$ 10 deg, and $<\theta_{orb}> \sim$ 7 deg, for the MWTD, and
thick disk, respectively.

\item The \enquote{coarse-grained} distribution of halo stars in the ($E$, $L_z$, $I_3$) phase-space shows the
Michie-Bodenheimer type
functional dependence on $E$ and $I_3$, such as $f(E,I_3) \propto \left[ \exp( - E / \sigma^2 ) - 1 \right]
\exp( - I_3 / r_a^2 \sigma^2 )$, where $\sigma$ is the 1d velocity dispersion and $r_{a}$ is the radius beyond which
the velocity dispersion is anisotropic. In the metallicity intervals of $-1.8 <$ [Fe/H] $< -1.4$ and $-2.2 <$ [Fe/H] $< -1.8$,
respectively, we obtain $(\sigma, r_a) = (122~{\rm km~s}^{-1}, 4.0~{\rm kpc})$ and $(116~{\rm km~s}^{-1}, 4.3~{\rm kpc})$.
In the most metal-poor interval of [Fe/H] $< -2.2$, we have found $(\sigma, r_a) = (145~{\rm km~s}^{-1}, 4.3~{\rm kpc})$,
i.e., the distributions of very metal-poor stars are systematically more extended along $E$ and $I_3$ than those in the higher metallicity ranges.

\item The inner halo stellar population, described in \citet{carollo2007,carollo2010}, includes a combination of GE debris stars, low-energy in-situ stars,
and a significant number of metal-poor prograde stars missed by the high-velocity cuts selection of
halo stars. 

\item Several of the retrograde substructures found in recent literature represent the \enquote{fine-grained} elements of the outer halo
defined in \citet{carollo2007,carollo2010}, and it is likely that the outer halo is made of a superposition of substructures,
relics of past accretion events.

\item The fraction of very metal-poor stars with [Fe/H]$< -2.2$, comprising the outer halo, relative to more metal-rich
halo stars, increases discontinuously at \I $\simeq 2000$ \kpckms, or $\theta_{\rm orb} \simeq 50$~deg,
corresponding to $z \simeq  10$~kpc at the solar radius. 

\item The $L_z$ distribution of halo stars shows a notable truncation in the prograde high-$L_z$ side, at $L_z \sim 1500$
\kpckms. On the other hand, the negative $L_z$ side shows an extended tail, indicating an asymmetry in the $L_z$
distribution, and the presence of a significant number of stars with large retrograde motions.
The truncation of the $L_z$ distribution at 1500 \kpckms \ may be connected to the "trail" feature identified in
the \I \ vs. $L_z$ diagram, in the direction of the Sgr dSph, and in the same phase space. This connection implies
that the Sgr dSph galaxy could induce a long-range dynamical effect on local halo stars, and generate a distinct
kinematical patterns.

\item Candidate in-situ stars, selected with binding energies of $E < -1.5 \times 10^5$ km$^2$~s$^{-2}$,
have orbital motions confined inside the solar radius, and orbital angles, $5^\circ \lesssim \theta_{\rm orb} \lesssim 20^\circ-30^\circ$
($z_{\rm max} < 5$ kpc). Notably, these stars are spread out in the [$\alpha$/Fe] vs. [Fe/H] diagram, but the majority of them
have [Fe/H]$< -1$ (See Figure~\ref{fig: lowest_E}).

\end{itemize}

The criteria for the selection of halo stars given in Equation (3) and (4) can be used for sample stars residing at any spatial locations, while this is not the case for the frequently adopted
method of $|V_{\rm star} - V_{\rm LSR}| > V_{lim}$ km~s$^{-1}$, because of
the unknown spatial dependence of halo's velocity distribution. 


As an example, let's take a nearby star located at $(x,y,z) = (R_{\odot},0,0)$.
The condition, $L_z < 1500$ \kpckms, suggests $v_{\phi} \lesssim 180$ km~s$^{-1}$.
At this solar position, $L_x=0$ and $L_y = - R_{\odot} v_z$ for any values of $v_x$,
so that $I_3$ is simply written as $I_3 = \frac{1}{2} v_z^2 (R_{\odot}^2 + \Delta^2)$. Then the condition of
\I $> 1000$ \kpckms \ for halo stars implies $v_z (R_{\odot}^2 + \Delta^2)^{1/2} > 1000$ \kpckms,
and if $\Delta = 4$~kpc is adopted, we obtain $v_z \gtrsim 110$ km~s$^{-1}$. It is worth emphasizing
that the extreme values of $v_{\phi} = 180$ km~s$^{-1}$ and $v_z = 110$ km~s$^{-1}$ (while $v_x$ takes any values), can locate a star inside the high-velocity cut of
$|V_{\rm star} - V_{\rm LSR}| = 200$ km~s$^{-1}$. Thus, the simple high-velocity cut misses many halo stars
in prograde rotation, and leads to an incomplete picture of the stellar halo.
Also, in the case of an arbitrary position along the Galactic plane at $(x,y,z)=(x,0,0)$, we obtain 
$v_z > 1000 / (x^2 + \Delta^2)^{1/2}$ km~s$^{-1}$, therefore the condition on $v_z$ generally depends on the position, $x$.
Thus, our orbit-based selection criteria for halo stars, given in Equation~(\ref{eq: for_halo}), correspond to a spatially dependent cut in velocity space, in contrast to the constant velocity cut applied to stars at all locations.

Our understanding of the Milky Way's stellar halo based on the analysis of local stellar samples, like those currently
adopted, is yet incomplete, due to the limited sampling generated by the survey footprints (Figure~\ref{fig: footprints}). Such local samples
miss stars with orbits whose apocentric radii are much
below the solar radius \citep{sommer1990,chiba2000}, i.e., those having very low $E$, including the candidate in-situ stars identified in this analysis.
Also, local stellar samples miss stars with nearly circular orbits in the outer parts of the halo, i.e., those showing
tangentially anisotropic velocities with $\beta < 0$ \citep[e.g.,][]{kafle2012,bird2020}.
Furthermore, current local samples lack metal-poor stars with low-$e$ orbits near the Galactic plane,
which have been identified in \citet{chiba2000} and \citet{sestito2020}.

The limited chemical-abundance information in several regions of the stellar halo is also challenging the understanding of
its true nature. The planned large spectroscopic surveys, over wide areas of the halo with
WEAVE \citep{dalton2012} and SDSS-V \citep{kollmeier2017}, and over narrow but more distant halo regions with Subaru/PFS \citep{takada2014,tamura2016},
in the near future, will significantly expand and deepen our knowledge of the stellar halo, and thus, its formation process.
Moreover, large ground-based photometric surveys designed to search for very metal-poor stars, like the Pristine Survey
\citep{starkenburg2017}, and to measure proper motions of very distant halo stars using 8-10 m telescopes \citep{qiu2020}
will be essential, combined with further {\it Gaia} data releases. Such surveys will bring us closer to obtain a complete
picture of the chemo-dynamical structure of the Milky Way's halo.\\


\section*{Acknowledgments}
We thank the referee for the useful comments on the manuscript.
M.C acknowledges support from the JSPS and MEXT Grant-in-Aid for Scientific Research
(No. 17H01101,  18H04434 and 18H05437).

Funding for the SDSS and SDSS-II has been provided by the Alfred P. Sloan Foundation, the Participating Institutions, the National Science Foundation, the U.S. Department of Energy, the National Aeronautics and Space Administration, the Japanese Monbukagakusho, the Max Planck Society, and the Higher Education Funding Council for England. The SDSS Web Site is http://www.sdss.org/.

Funding for the Sloan Digital Sky Survey IV has been provided by the Alfred P. Sloan Foundation, the U.S. Department of Energy Office of Science, and the Participating Institutions. SDSS acknowledges support and resources from the Center for High-Performance Computing at the University of Utah.

The SDSS is managed by the Astrophysical Research Consortium for the Participating Institutions. The Participating Institutions are the American Museum of Natural History, Astrophysical Institute Potsdam, University of Basel, University of Cambridge, Case Western Reserve University, University of Chicago, Drexel University, Fermilab, the Institute for Advanced Study, the Japan Participation Group, Johns Hopkins University, the Joint Institute for Nuclear Astrophysics, the Kavli Institute for Particle Astrophysics and Cosmology, the Korean Scientist Group, the Chinese Academy of Sciences (LAMOST), Los Alamos National Laboratory, the Max-Planck-Institute for Astronomy (MPIA), the Max-Planck-Institute for Astrophysics (MPA), New Mexico State University, Ohio State University, University of Pittsburgh, University of Portsmouth, Princeton University, the United States Naval Observatory, and the University of Washington.

{}


\appendix


\section{Mass model and orbital motion}
\label{section: appendix_A}

Here, we describe our adopted Galactic potential of St\"ackel type based on the model originally proposed by \citet{sommer1990}. The complete description of a St\"ackel potential is given in \citet{dezeeuw1985} and \citet{dejonghe1988}.

\begin{figure}[t!]
\centering
\includegraphics[width=75mm]{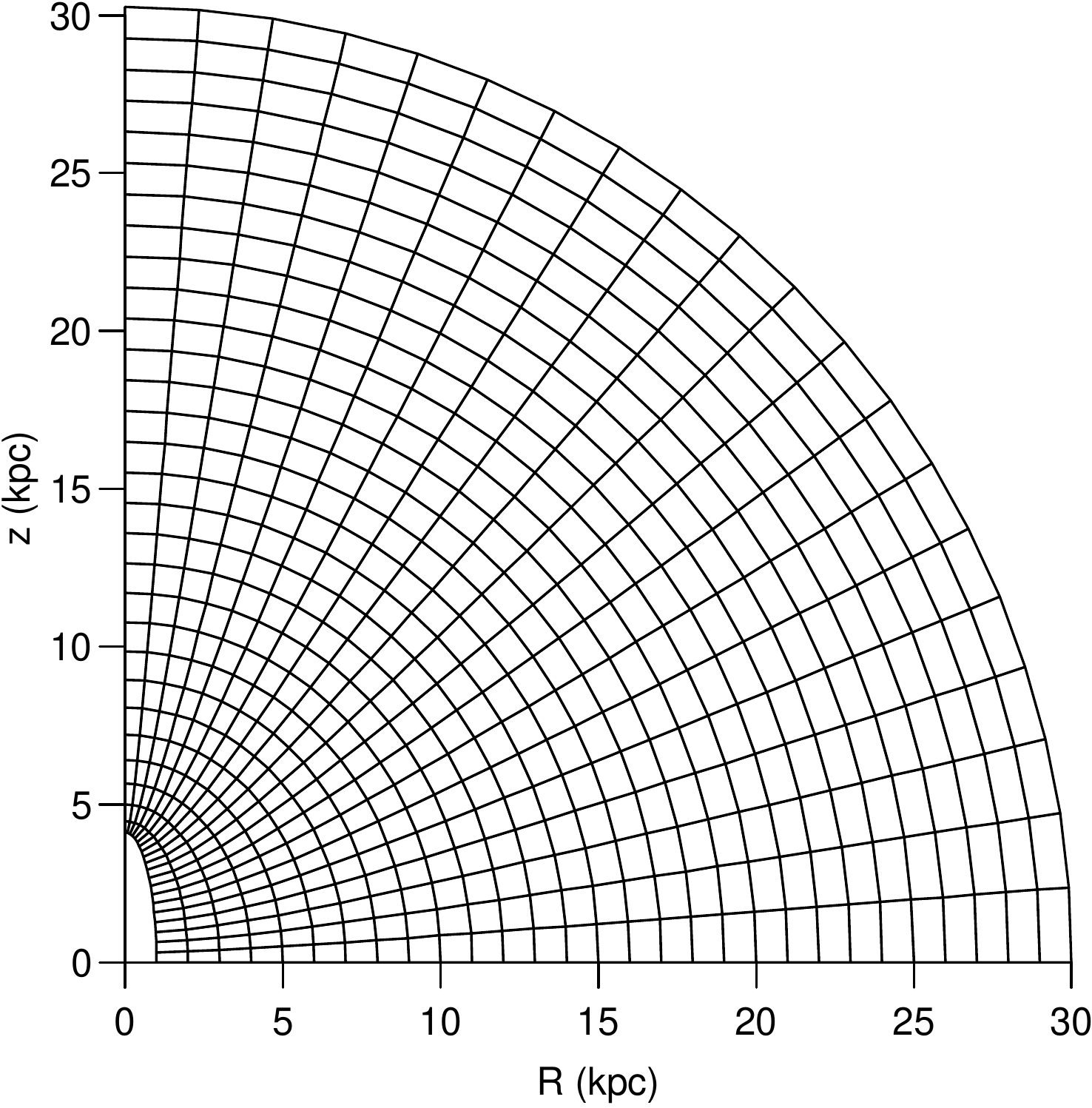}
\caption{
The grids of the spheroidal coordinates $(\lambda, \nu)$ in the meridional plane, defined with the focus
along the $z$ axis at $\Delta = 4$ kpc.
The orbits in an axisymmetric gravitational potential of St\"ackel type
are bounded by $\lambda = {\rm const.}$ and $\nu = {\rm const.}$, the surfaces of a prolate and hyperboloid, respectively.
}
\label{fig: coord}
\end{figure}

We adopt the axisymmetric potential defined in spheroidal coordinates $(\lambda,\phi,\nu)$, where $\phi$ is
the azimuthal angle in cylindrical coordinates of $(R, \phi, z)$ and $(\lambda,\nu)$ are the roots for $\tau$ of
\begin{equation}
    \frac{R^2}{\tau + \alpha} + \frac{z^2}{\tau + \gamma} = 1 \ ,
\end{equation}
where $\alpha$ and $\gamma$ are constants, constraining the range of $\lambda$ and $\nu$ as
$-\gamma \le \nu \le -\alpha \le \lambda$. The coordinates are defined with the focal distance
$\Delta = \sqrt{\gamma - \alpha}$ along the $z$ axis and their grids are displayed in Figure~\ref{fig: coord}.
The relation between $(\lambda, \nu)$ and $(R, z)$ is then written as
\begin{equation}
    R^2 = \frac{(\lambda+\alpha)(\nu+\alpha)}{\alpha-\gamma}, \quad
    z^2 = \frac{(\lambda+\gamma)(\nu+\gamma)}{\gamma-\alpha} \ .
\end{equation}

The gravitational potential of this type, $\psi(\lambda, \nu)$, is generally given as
\begin{equation}
\psi(\lambda, \nu) = - \frac{(\lambda+\gamma)G(\lambda)-(\nu+\gamma)G(\nu)}{\lambda - \nu} \ , 
\end{equation}
where $G(\tau)$ with $\tau = \lambda, \nu$ is an arbitrary function. In this work, the Galactic potential is represented
by the combination of a disk, $G_{\rm D}(\tau)$, and a dark halo $G_{\rm H}(\tau)$, thus
given as $G(\tau) = G_{\rm D}(\tau) + G_{\rm H}(\tau)$.
Following Sommer-Larsen \& Zhen (1990), the density distribution of a disk is given by a highly flattened, perfect oblate
spheroid, resembling a Kuzmin disk, and that of a dark halo is modelled with a slightly flattened, oblate spheroid,
resembling an isothermal sphere, originally developed by de~Zeeuw, Peletier \& Franx (1986) (their $s=2$ model).
Then, $G_{\rm D}$ and $G_{\rm H}$, respectively, are written as (with the gravitational constant, $G_{\rm gr}$)
\begin{eqnarray}
G_{\rm D}(\tau) &=& \frac{2G_{\rm gr}}{\pi} \frac{M_{\rm D}}{\sqrt{\tau+\gamma}}  \arctan \sqrt{\frac{\tau+\gamma}{-\gamma}} \\
G_{\rm H}(\tau) &=& -4\pi G_{\rm gr} \rho_0 (\gamma-b) \nonumber \\
                &\times& \left[ \ln \frac{\Delta^2-\gamma+b}{-\gamma+b}-\frac{\tau+\gamma+\Delta^2}{2(\tau+\gamma)} \ln\frac{\tau+b}{-\gamma+b}
                 + \frac{\Delta^2+\gamma-b}{\sqrt{-\gamma+b}}
                    \left( \frac{1}{\sqrt{\tau+\gamma}} \arctan\sqrt{\frac{\tau+\gamma}{-\gamma+b}} 
                    - \frac{1}{\Delta} \arctan \frac{\Delta}{\sqrt{-\gamma+b}} \right) \right] \ ,
\end{eqnarray}
where $M_{\rm D}$ is the mass of a disk and $\rho_0$ and $b$ are parameters for a dark halo\footnote{There is a typo in equation (9) for $G_{\rm H}$ in the Sommer-Larsen \& Zhen (1990) paper, but their calculations adopted the correct one as given here.}.
These parameters are chosen so as 
to reproduce the observed constraints for the Galactic potential. Here we adopt the set of these parameters in Sommer-Larsen
\& Zhen (1990), where the rotation curve is flat beyond $R \simeq 4$~kpc with a rotation velocity of $223$ km~s$^{-1}$ at
the position of the Sun $R = 8.5$~kpc (see their Figure 2): $\Delta = \sqrt{\gamma-\alpha} = 4$~kpc, $\sqrt{-\gamma}=0.125$~kpc,
$\sqrt{b-\gamma}=6$~kpc, $M_{\rm D} = 9.0 \times 10^{10}$~$M_{\odot}$ and $\rho_0 = 2.45 \times 10^7$~$M_{\odot}$~kpc$^{-3}$.
We further set $G=G_{\rm D}+G_{\rm H}=0$ at $R_{\rm t} = 200$~kpc in the Galactic plane, i.e.,
at $\lambda = R_{\rm t}^2 - \alpha$, by adding the corresponding constant to $G$, thereby giving $\psi = 0$ at this kind of tidal radius. This adjustment in $G$ is required to bound the sample stars in this potential.

The orbital motion in this potential has the Hamiltonian $H$,
\begin{equation}
    H = \frac{p_\lambda^2}{2P^2} + \frac{p_\phi^2}{2Q^2} + \frac{p_\nu^2}{2R^2} + \psi(\lambda, \nu) \ , 
\end{equation}
with
\begin{equation}
    P^2 = \frac{\lambda-\nu}{4(\lambda+\alpha)(\lambda+\gamma)},
    \quad Q^2 = - \frac{\lambda-\nu}{4(\nu+\alpha)(\nu+\gamma)} \ .
\end{equation}
The momentum components conjugate to $(\lambda, \phi, \nu)$, $(p_\lambda,p_\phi,p_\nu)$, are given as
\begin{equation}
    p_\lambda = P^2 \dot{\lambda}, p_\phi = R^2 \dot{\phi}, p_\nu = Q^2 \dot{\nu} \ ,
\end{equation}
where $\dot{}$ denotes the time derivative and $p_\phi = L_z$.  The equations of motion are then written as
\begin{eqnarray}
    p_\tau^2 &=& \frac{1}{2(\tau+\alpha)}
    \left[ G(\tau) - \frac{I_2}{\tau+\alpha} - \frac{I_3}{\tau+\gamma} - |E| \right], \quad \tau=\lambda,\nu 
    \label{eq: momentum} \\
    p_\phi^2 &=& L_z^2 = 2I_2 \ ,
\end{eqnarray}
where $E(=H)$, $I_2=L_z^2/2$ and $I_3$ are the three isolating integrals of motion. $I_3$ is explicitly written as
\begin{equation}
I_3 = \frac{1}{2}(L_x^2 + L_y^2) + \Delta^2 \left[ \frac{1}{2}v_z^2 - z^2 \frac{G(\lambda)-G(\nu)}{\lambda - \nu} \right] \ ,
\end{equation}
where $L_x$ and $L_y$ are the angular momentum components in the $x$ and $y$ directions, respectively, and $v_z$ is
the velocity component in the $z$ direction.

\section{Description of the data}

\subsection{SDSS DR7 Calibration Stars}
\label{section: appendix_B1}

SDSS is now arrived to its fourth generation and its 16th data release (\citealt{ahumada2019}, and reference therein).\\
The SDSS and its extensions uses a dedicated 2.5 m
telescope \citep{gunn2006} located at the Apache Point
Observatory in New Mexico. The telescope is equipped with
an imaging camera and a pair of spectrographs, each of which
are capable of simultaneously collecting 320 medium-resolution
(R $\sim$ 2000) spectra over its 7 deg2 field of view, so that on the
order of 600 individual target spectra and, roughly, 40 calibration star
and sky spectra are obtained on a given spectroscopic “plugplate”
\citep{york2000}. The SEGUE sub-survey, carried out as part of SDSS-II, ran
from 2005 July to 2008 July. SEGUE obtained some 240,000
medium-resolution spectra of stars in the Galaxy, selected to explore
the nature of stellar populations from 0.5 kpc to 100 kpc
\citep{yanny2009}. SDSS and SEGUE obtained spectroscopy and photometry for
a small number (16) of calibration stars obtained for each
spectroscopic plug plate, chosen for two primary reasons. The
first set of these objects, the spectrophotometric calibration stars,
includes stars that are selected to approximately remove the distortions
of the observed flux of stars and galaxies arising from
the wavelength response of the Astrophysical Research Consortium
(ARC) 2.5 m telescope and the SDSS spectrographs,
as well as the distortions imposed on the observed spectra by
Earth’s atmosphere. The spectrophotometric calibration stars
cover the apparent magnitude range 15.5 $<$ g0 $<$ 17.0, and satisfy
the color ranges 0.6 $<$ (u - g)$_{0}$ $<$ 1.2; 0.0 $<$ (g - r)$_{0}$ $<$
0.6. The second set of stars, the telluric calibration stars, is used
to calibrate and remove night-sky emission and absorption features
from SDSS spectra. The telluric calibration stars cover the
same color ranges as the spectrophotometric calibration stars,
but at fainter apparent magnitudes, in the range 17.0 $<$ g$_{0}$ $<$
18.5. Stellar parameters for the calibration stars were obtained by employing the SEGUE Stellar Parameter Pipeline (SSPP; \citealt{lee2008a,lee2008b,lee2011a}) which processes the
wavelength- and flux-calibrated spectra generated by the standard
SDSS spectroscopic reduction pipeline, obtains equivalent
widths and/or line indices for about 80 atomic or molecular
absorption lines, and estimates the effective temperature, Teff ,
surface gravity, log g, and metallicity, [Fe/H], for a given star
through the application various approaches. 
The internal errors for stars in the temperature range
that applies to the calibration stars are $\sigma_{\rm Teff} \sim$ 100 K to $\sim$ 125 K, $\sigma_{\rm log g} \sim $ 0.25 dex, and $\sigma_{\rm [Fe/H]} \sim$ 0.20 dex. 

The SSPP pipeline derives also the $\alpha$-elements abundances which are available in the SDSS Catalog Archive Server or CAS \citep{thakar2008}, only for the SDSS-DR7. The pipeline adopts 
of a pre-existing grid of synthetic spectra \citep[NEWODF;][]{castelli2003}, with no 
enhancement in $\alpha$-element abundances, and creates a fine (steps of 0.2 dex for 
$\log g$ and 0.2 dex for [Fe/H]) grid of spectra by interpolation between the wider model 
grids (steps of 0.5 dex). The wavelength range is 4500 $-$ 5500 $\AA$, chosen because 
it contains a large set of metallic lines, but avoids the CH $G$-band feature 
($\sim$ 4300 $\AA$, which can be strong in metal-poor stars) and the Ca II K 
( $\sim$ 3933 \AA) and H ($\sim$ 3968 $\AA$) lines, which can saturate for cool 
metal-rich stars. The final grid covers 4000 K $<$ $T_{\rm eff}$  $<$ 8000 K, in steps 
of 250 K, 0.0 $<$ $\log g$ $<$ 5.0, in steps of 0.2 dex, and $-$4.0 $<$ [Fe/H] $< -$0.4, 
in steps of 0.2 dex. The range in [$\alpha$/Fe] introduced for the spectral synthesis 
covers +0.1 $<$ [$\alpha$/Fe] $<$ +0.6, in steps of 0.1 dex, at each node of $T_{\rm eff}$, 
$\log g$, and [Fe/H]. After creation of the full set of synthetic spectra, they are 
degraded to SEGUE resolution (R = 2000) and re-sampled to 1 $\AA$ wide linear pixels 
(during SSPP processing, the SEGUE spectra are also linearly re-binned to 1 $\AA$ per 
pixel). In the SSPP pipeline, the notation [$\alpha$/Fe] denotes an average of the 
abundance ratios for individual $\alpha$-elements weighted by their line strengths 
in synthesized spectra. In the selected spectral range the dominant features are the 
magnesium (Mg) and titanium (Ti) lines, which are the primary contributors to the 
determination of [$\alpha$/Fe], with some influence from silicon (Si) and calcium 
(Ca). In the adopted wavelength range, and at the SDSS spectral resolution, oxygen 
(O) has no strong detectable features, and it is excluded in the computation of the 
overall $\alpha$-element abundance. The [$\alpha$/Fe] measurements were validated 
with the stars in other external sources such as the ELODIE (\citealt{prugniel2001}, and later releases) 
spectral library, and compared with those obtained by analyzing a large sample of 
SEGUE stars observed at high spectral resolution. The SSPP provides [$\alpha$/Fe] 
abundance for SDSS/SEGUE spectra with a precision of $\sim$ 0.06 dex at S/N $>$ 50 
and $<$ 0.1 dex at S/N = 20. A detailed description can be found in \citet{lee2011a}.\\
Radial velocities for stars in our sample are derived from matches to an external 
library of high-resolution spectral templates with accurately known velocities 
\citep{allende2007,yanny2009}, degraded in resolution to that of the SDSS 
spectra. The typical precision of the resulting radial velocities are on the order 
of 3-20 km s$^{-1}$, depending on the S/N of the spectra, with zero-point errors of no 
more than 3 km s$^{-1}$, based on a comparison of the subset of stars in our sample 
with radial velocities obtained from the high-resolution spectra taken for testing 
and validation of the SSPP.
The initial sample employed in this analysis consists of $\sim$ 32,000 unique stars 
selected in the temperature range 4500 K $<$ $T_{\rm eff}$ $<$ 7000 K, where the SSPP 
pipeline provides the highest accuracy for the derived atmospheric parameters. 

\subsubsection{SDSS DR7 full sample}
\label{section: appendix_B1.1}

In the SDSS archive we selected only stars with high signal-to-noise (SNR $>$ 40), and with critical and cautionary flags provided by the SSPP pipeline set to {\it normal}. Stars are selected in the temperature range of 4500 $^{\circ}$ $<$ T$_{eff} <$ 7000 $^{\circ}$ where the SSPP pipeline provides the best estimate for the stellar parameters and abundances. This selection provides N = 65,678 stars.    

\subsection{SDSS DR16}
\label{section: appendix_B2}

In this paper we also make use of data from Data Release 16 of the Apache Point Observatory Galactic Evolution Experiments (APOGEE; \citealt{majewski2017}). The APOGEE survey is obtaining high resolution spectroscopy across the entire Milky Way Galaxy with two near-infrared, high-resolution multiplexed spectrographs, the 2.5m Sloan Foundation Telescope at APO, in the northern hemisphere, and the 2.5m Irene du Pont Telescope at LCO (Las Campanas Observatory), in the southern hemisphere. DR16 \citep{ahumada2019} is the fourth data release of SDSS-IV \citep{eisenstein2011}, containing all SDSS-III APOGEE-1 data and SDSS-IV APOGEE-2 data acquired with both instruments through August
2018. DR16 contains information for 437,485 unique stars, including reduced spectra, radial velocities, atmospheric parameters, and individual elements abundances for 15 chemical elements with the ASPCAP pipeline (C, N, O, Na,Mg, Al, Si, S, K, Ca, Ti, V, Mn, Fe, and Ni), \citep{holtzman2015,zamora2015,garciaperez2016}). We refer the reader to \citep{holtzman2018,holtzman2018,jonsson2020} for a complete and detailed description of the APOGEE data, namely DR14 and DR16. We selected the APOGEE stellar sample from the CASJobs archive service by using the flags provided in the data files, and described in \citet{holtzman2015}. In particular, we rejected stars in the database with unreliable stellar parameters and abundances, such that, VERY$\_$BRIGHT$\_$NEIGHBOR, PERSIST$\_$HIGH, and SNR$\_$BAD are set in STARFLAGS, and STAR$\_$BAD, METAL$\_$WARN, ROTATION$\_$WARN, and METALS$\_$BAD are set in the ASPCAPFLAGS. In addition, we applied other cuts in T$_{\rm eff}$ and log(g) in order to avoid some issues described in \citep{holtzman2015}, and we only consider stars with T$_{\rm eff}$ $>$ 4000 K, and 1.0 $<$ log(g) $<$ 3.5. The restriction in T$_{\rm eff}$ is due to the low quality of the ASPCAP fitting at cooler temperatures, while that in log(g) is applied because stars with log(g) $\ge$ 4.0 have strong deviation from asteroseismic gravities \citep{holtzman2015}. Finally, we selected stars with signal-to-noise, S/N $>$ 80. By applying the Apogee flags, stellar parameters cuts, and S/N selection we obtained a subsample of N = 70,130 stars. 
In our analysis we have considered the calibrated [M/H] as the
indicator of the global metallicity, because it is corrected for the
overestimation on the metal-poor side, while for the [$\alpha$/Fe] indicator we choose [Mg/Fe]. Typical errors these abundances are, $\sigma \sim$ 0.02-0.04.
Note that it is advisable the use of [Fe/H] and [Mg/Fe] because some suspicious values have been removed from these quantities, and there exist also a curious thin $\alpha$-finger in the [$\alpha$/M] vs. [M/H] diagram that is not present at all in the [Mg/Fe] vs. [Fe/H] diagram \citep{jonsson2020}.   

\begin{figure*}[hbt!]
\vspace{-4cm}
\includegraphics[scale=0.7,angle=90]{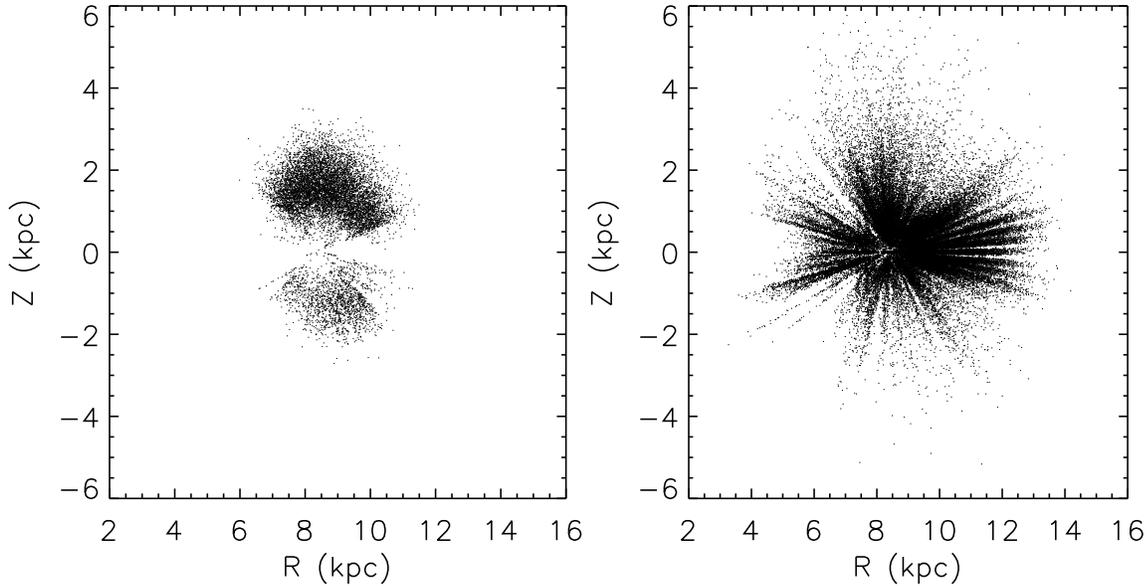}
\caption{
The distribution of the SDSS DR7 calibration stars (left panel), and the SDSS DR16 APOGEE subsamples (right panel), in the
($R,z$) plane, where $R$ is the projected Galactocentric distance, and $z$ is the distance from the Galactic plane. 
}
\label{fig: footprints}
\end{figure*}

\subsection{Distances, proper motions, and footprints of these catalogs}
\label{section: appendix_B3}

The SDSS DR7 calibration stars, the DR7 full sample, and APOGEE DR16 were
cross-matched with the Gaia DR2 database \citep{brown2018} to retrieve accurate astrometric 
positions, trigonometric parallaxes, and proper motions, using the CDS (Centre de 
Données Astronomiques de Strasbourg) X-Match service, and adopting a very small search 
radius (0.8'' for DR7 and 0.6'' for APOGEE) to avoid duplicates. The match provides positions, parallaxes, and 
proper motions for all of the stars in the sample. We select stars with relative 
parallax errors of $\sigma_{\pi}/\pi <$ 0.2, and derive their distance estimates 
using the relation $d = 1/\pi$ (and used these distances to select stars with 
heliocentric distance d $\leq$ 4 kpc). This selection reduces the number of stars 
to 10,820 for the DR7 calibration stars ($\sim$ 1/3rd of the initial sample), and 62,060 for the Apogee sample. The majority of stars in this final 
sample have errors on proper motions below 0.2 mas yr$^{-1}$. We also select stars 
with Galactocentric distance in the range 6 kpc $<$ $R$ $<$ 12 kpc to include extended regions of the 
galactic disk. The final number of stars after this selection is 10,814, and 58,912, for DR7 calibration stars and APOGEE DR16, respectively. 
In case of the full SDSS SEGUE DR7 sample, the same selection provides a final number of stars of N = 46,500.


In this analysis the parallaxes have been corrected for the parallax zero-point offset \citet{lindegren2018}.  In \citet{carollo2019} was established that the addiction of a constant offset model improves the agreement between the distances derived from the inverse of parallax with those inferred with the Bailer-Jones's method. Therefore, we have corrected the parallaxes with a constant value of 0.054 mas which can be justified by following the results reported in \citet{leung2019}, where it is shown that the constant offset model is in very good agreement with the multi-variate offset model up to 10 kpc for a sample of $\sim$ 265,000 stars in common between APOGEE DR14 \citep{holtzman2015,garciaperez2016,abolfathi2018} and {\it Gaia} DR2.

\subsection{The phase space distributions of the DR7 full sample and DR16}
\label{section: appendix_B4}

\begin{figure*}[hbt!]
\centering
\vspace*{-0.4cm}
\includegraphics[width=170mm]{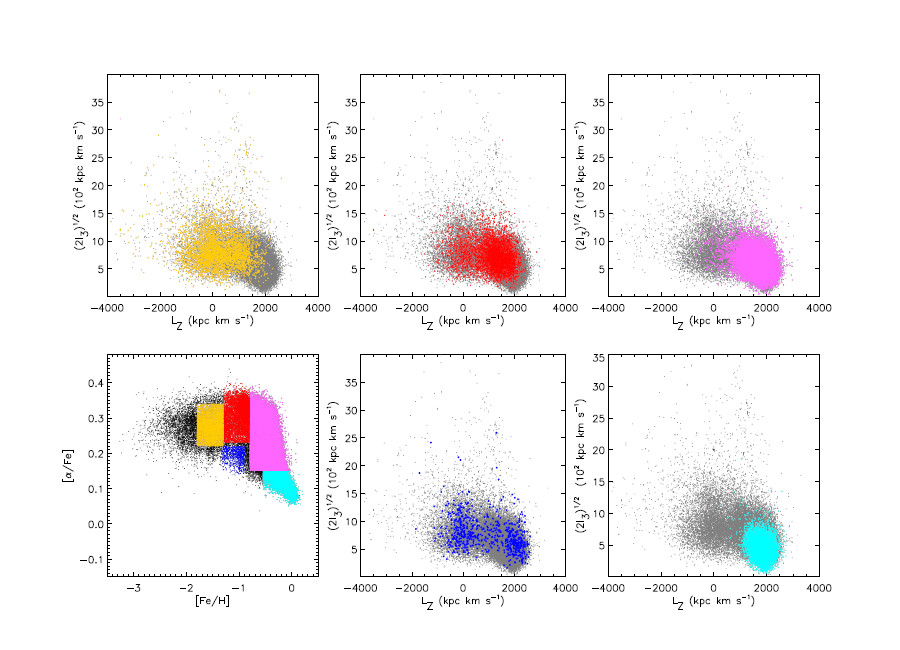}
\caption{
The same as Figure~\ref{fig: phase1_sub} but for the SDSS DR7 full sample.
}
\label{fig: DR7full_phase1_sub}
\end{figure*}

\begin{figure*}[hbt!]
\centering
\vspace*{-0.4cm}
\includegraphics[width=160mm]{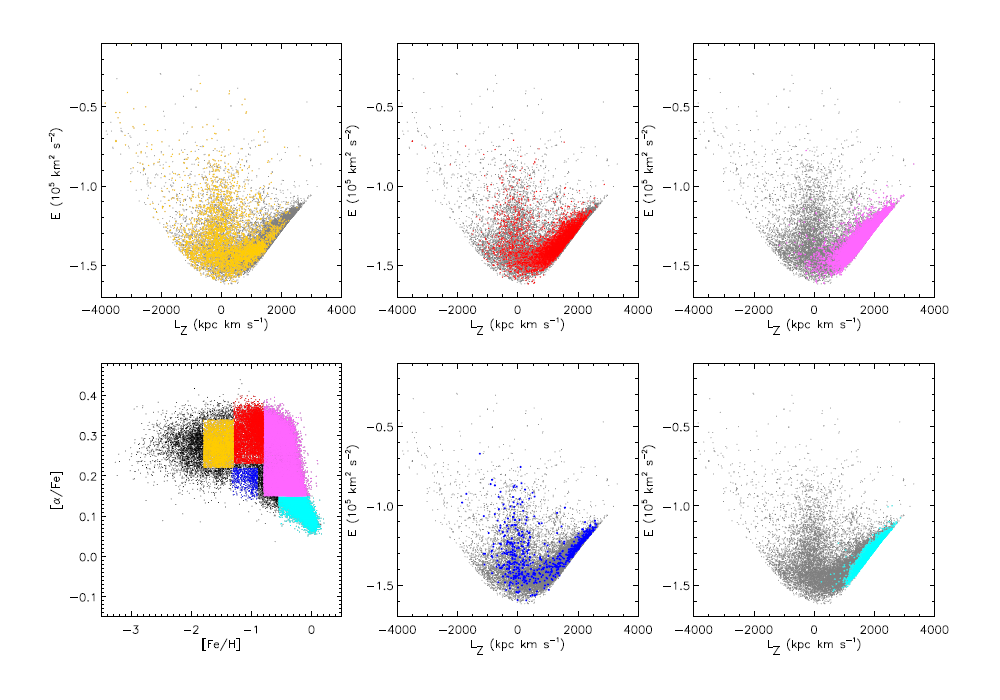}
\caption{
The same as Figure~\ref{fig: phase2_sub} but for the SDSS DR7 full sample.
}
\label{fig: DR7full_phase2_sub}
\end{figure*}

\begin{figure*}[hbt!]
\centering
\vspace*{-0.4cm}
\includegraphics[width=140mm]{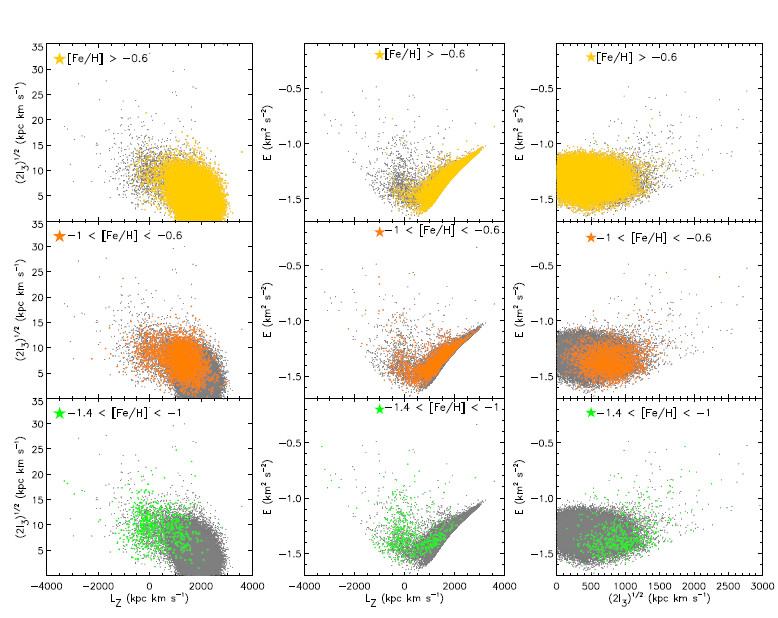}
\caption{
The same as Figure~\ref{fig: phase1} but for the SDSS DR16 APOGEE stars.
}
\label{fig: DR16_phase1}
\end{figure*}

\begin{figure*}[hbt!]
\centering
\vspace*{-0.4cm}
\includegraphics[width=140mm]{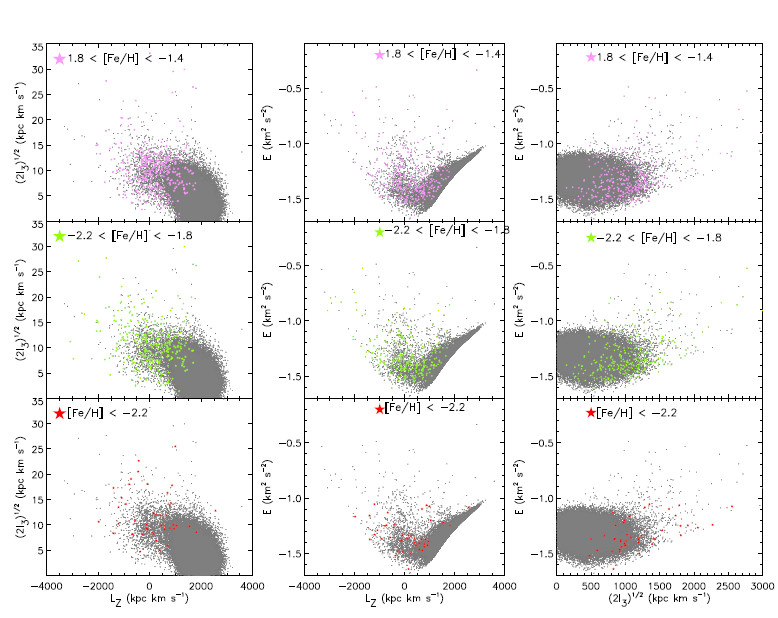}
\caption{
The same as Figure~\ref{fig: phase1} but for the SDSS DR16 APOGEE stars.
}
\label{fig: DR16_phase2}
\end{figure*}

In Section 3 of the main paper, we present the phase-space distributions of the SDSS DR7 calibration stars for each interval of [Fe/H], as well as for the selected box areas in the [$\alpha$/Fe] vs. [Fe/H] diagram. We also show the phase-space distributions in these [$\alpha$/Fe] vs. [Fe/H] box areas for the SDSS DR16.
Here, for the sake of comparison and also completeness, we show the phase-space distributions for the SDSS DR7 full sample, in the same box areas (Figure~\ref{fig: DR7full_phase1_sub} and
\ref{fig: DR7full_phase2_sub}), while in Figure~\ref{fig: DR16_phase1} and
\ref{fig: DR16_phase2} it is shown the SDSS DR16 APOGEE phase-space as a function of the metallicity.\\
The SDSS DR7 full sample shows basically the same phase-space distributions as the DR7 calibration stars sample, and actually,
enhances the main properties found in section 3.1 of the main paper, because of the larger number of stars in each bin of metallicity.
The phase-space distribution of the DR16 Apogee differs with respect that of the DR7 samples, because of the footprints (Figure~\ref{fig: footprints}). In particular, DR16 has a larger coverage in $R$
($3 \lesssim R \lesssim 14$ kpc), which lead to a broader distribution of stars around $L_z \sim 2000$ \kpckms \ and \I $\sim 0$ \kpckms, in the metal-rich range, [Fe/H] $> -$0.6. Also, at $L_z \simeq 0$ \kpckms, the DR16 APOGEE stars show somewhat lower energy values than the DR7 stellar samples. This is likely due to the fact that some stars in APOGEE are located at smaller $R$, and thus have smaller binding energies than DR7 stars.


\end{document}